\documentclass[sigconf]{acmart}

\AtBeginDocument{%
  \providecommand\BibTeX{{%
    \normalfont B\kern-0.5em{\scshape i\kern-0.25em b}\kern-0.8em\TeX}}}

\copyrightyear{2024}
\acmYear{2024}
\setcopyright{rightsretained}
\acmConference[CHI '24]{Proceedings of the CHI Conference on Human Factors in Computing Systems}{May 11--16, 2024}{Honolulu, HI, USA}
\acmBooktitle{Proceedings of the CHI Conference on Human Factors in Computing Systems (CHI '24), May 11--16, 2024, Honolulu, HI, USA}
\acmDOI{10.1145/3613904.3642120}
\acmISBN{979-8-4007-0330-0/24/05}

\usepackage{graphicx}  
\usepackage{caption}
\usepackage{subcaption}
\usepackage[finalnew]{trackchanges}

\begin{document}

\title[Mapping the Design Space of Teachable Social Media Feed Experiences]{Mapping the Design Space of \\Teachable Social Media Feed Experiences}

\author{K. J. Kevin Feng}
\affiliation{%
  \institution{University of Washington}
  \city{Seattle}
  \country{USA}}
\email{kjfeng@uw.edu}

\author{Xander Koo}
\affiliation{%
  \institution{Georgia Institute of Technology}
  \city{Atlanta}
  \country{USA}}
\email{xander@gatech.edu}

\author{Lawrence Tan}
\affiliation{%
  \institution{University of Washington}
  \city{Seattle}
  \country{USA}}
\email{lawtan@cs.uw.edu}

\author{Amy Bruckman}
\authornote{Equal contribution as senior authors.}
\affiliation{%
  \institution{Georgia Institute of Technology}
  \city{Atlanta}
  \country{USA}}
\email{asb@cc.gatech.edu}

\author{David W. McDonald}
\authornotemark[1]
\affiliation{%
  \institution{University of Washington}
  \city{Seattle}
  \country{USA}}
\email{dwmc@uw.edu}

\author{Amy X. Zhang}
\authornotemark[1]
\affiliation{%
  \institution{University of Washington}
  \city{Seattle}
  \country{USA}}
\email{axz@cs.uw.edu}

\renewcommand{\shortauthors}{Feng, et al.}

\begin{abstract}
  Social media feeds are deeply personal spaces that reflect individual values and preferences. However, top-down, platform-wide content algorithms can reduce users' sense of agency and fail to account for nuanced experiences and values.
  %, and marginalize communities with unique values and customs.
  Drawing on the paradigm of interactive machine teaching (IMT), an interaction framework for non-expert algorithmic adaptation, we map out a design space for \textit{teachable social media feed experiences} to empower agential, personalized feed curation. To do so, we conducted a think-aloud study ($N=24$) featuring four social media platforms---Instagram, Mastodon, TikTok, and Twitter---to understand key signals users leveraged to determine the value of a post in their feed. We synthesized users' signals into taxonomies that, when combined with user interviews, inform five design principles that extend IMT into the social media setting. We finally embodied our principles into three feed designs that we present as sensitizing concepts for teachable feed experiences moving forward. 
\end{abstract}




\maketitle

\section{Introduction}
\label{s:intro}
Social media feeds shape our everyday online interactions with others. Their interface designs and affordances define boundaries on \textit{how} we can engage with people and content, while their algorithms dictate both \textit{what} and \textit{who} we engage with in the first place. Taken together, they shape our collective behaviors and social norms, cementing their role as social architects in the digital age.

The rise of algorithmic content curation and distribution via feeds happened alongside a key shift in the ownership structure of social networks---in the past 20 years, social media platforms transitioned away from being hosted by distributed, independent servers to operating under a small number of private corporations reaching millions of people worldwide \cite{denardis2015internet}. The funneling of capacities for algorithmic curation into the hands of a select few results in what Reviglio et al. call a lack of ``algorithmic sovereignty'' \cite{reviglio2020thinking}. It is by this process that the complexity and richness of our social realities have been distilled into a small number of homogenized parameters. In the face of seemingly ubiquitous promises of in-feed personalization, this has brought about a ``personalization paradox'' \cite{simpson2022tame}. \looseness=-1

Increasingly centralized curation can have significant negative consequences for users' agency \cite{Kim2021TeachingLearningIA, sap2019risk, jia2022misinfo, pavlov2014user, lukoff2021agency, baughan2022dissociation}. Platforms today often employ a top-down, one-size-fits-all approach when it comes to platform design and governance. In doing so, they marginalize those who fail to conform to the platform-wide majority. For example, user groups with culturally significant patterns of language use may have their content mislabeled and ``downranked'' by platforms \cite{sap2019risk, jia2022misinfo}, while neurodiverse users may find many posts too overwhelming to consume \cite{pavlov2014user}. Even users who fit within the majority may get frustrated from being shown irrelevant content with no efficient way to set controls that would eliminate such content from their feeds. 

Prior work has documented users' attempts to reclaim their agency by deriving algorithmic folk theories to probe black-box feed curation algorithms \cite{eslami2015assumed, eslami2016folk, devito2018folk, lee2022tiktok, karizat2021tiktok, siles2020folk} and ``teaching'' these algorithms to better align with their preferences through strategic in-feed interactions \cite{Kim2023InvestigatingHU, eslami2016folk}. The efficacy of these ad-hoc techniques, however, is often unclear, and using them can even leave users with undesirable feelings of coercion and manipulation \cite{burrell2019twitter}. Why might this be? The problem is unlikely to be rooted in the \textit{quantity of feedback} that the user provides to the algorithm---after all, modern recommender systems leverage a wide variety of both implicit (e.g., content dwell time, mouse movements) and explicit (e.g., likes, blocks) feedback elicitation techniques \cite{Stray2022BuildingHV, narayanan-algorithms, Knijnenburg2011EachTH, Mendez2017UserGA}, sometimes learning user preferences to a startling degree of accuracy \cite{simpson2022tame}. Instead, we posit that this is due to users' lack of opportunity to \textit{agentially articulate pertinent feedback} to the algorithm, and have the algorithm respond accordingly based on their feedback. 

Given this, we draw upon literature in interactive machine teaching (IMT) \cite{ramos2020interactive, ng2020knowledge, wall2019using} to chart out avenues for enabling \textit{teachable feed experiences\footnote{By ``experiences,'' we refer to the combination of in-feed interface affordances and underlying algorithmic behavior orchestrated by, or enabled using, those affordances.}} on social media. IMT proposes an interaction framework by which a user (the human teacher) without expertise in machine learning can train a model (the algorithmic learner) to accomplish desired tasks with limited amounts of pre-labelled data. Applications of IMT in social media settings, however, has been limited. To extend IMT's framework to social media, we conducted a think-aloud study ($N=24$) with users from four feed-based social media platforms with diverse cultures and affordances---Instagram, Mastodon, TikTok, and Twitter\footnote{As of July 23, 2023, Twitter has been rebranded to X. Since the platform was still known as Twitter during the study, we will refer to it as so throughout the paper.}---to answer the following question: 

\begin{quote}
    \textbf{What are prominent signals relied upon by users to judge the value of content in their feeds, and are thus amenable to teaching to an algorithmic learner? } 
\end{quote}

We define a ``signal'' as a pair of one feature (a category of information that can be extracted from a post, such as the \textsc{author} or \textsc{included hashtags}) and one characteristic (a statement describing the significance of the feature, such as \textit{``is part of a recent fashion trend I've been following''} for the feature \textsc{included hashtags})\footnote{For more on features and characteristics, see Section \ref{s:study-procedure}.}. We define ``value'' broadly based on the \textit{desired order of consumption}: Post A has a higher value than Post B if the user prefers to see A before B in their feed.

We find from our study that users leveraged a variety of signals to evaluate content from their feeds. These evaluations are  nuanced in ways that current preference elicitation methods, such as ``likes'' or content dwell time, may fail to capture. Many users also expressed desires for feed experiences centered around individuals they cared about rather than content-based recommendations, better management of saved content, and more agency---in particular self-causality---when curating their feeds. Supplementing our findings with prior work on IMT, we offer five IMT-inspired design principles for teachable feeds. Finally, we embody these principles into three proposed feed designs that serve as sensitizing concepts to catalyze future research in this area. 

Concretely, our paper offers the following contributions:

\begin{enumerate}
    \item Cross-platform taxonomies of prominent signals used to determine the value of posts in social media feeds, enriched by themes extracted from user interviews.
    \item Five principles to guide the design of teachable social media feed experiences.
    \item Three proposed feed designs that illustrate our principles and serve as sensitizing concepts for teachable social media feed experiences going forward.
\end{enumerate}

These contributions pave a path to equipping today's social media systems with novel design patterns to empower agential, personalized feed curation.

\section{Related Work}

To motivate our work, we review prior literature on agency and algorithms on social media, incorporating human values into content distribution algorithms (most notably recommender systems), and interactive machine teaching.

\subsection{Agency and Algorithms on Social Media}
In HCI literature, increasing user agency is often discussed as an aspirational ideal \cite{bennett2023agency, lukoff2021agency, zhang2022screen, baughan2022dissociation, coyle2012agency, madary2022illusion, purohit2023diets}. But how exactly do HCI researchers understand agency? Bennett et al. \cite{bennett2023agency} surveyed 161 publications across 30+ years of HCI and identified 4 key aspects of agency: self-causality/identity, material/experiential, interaction time-scales, and tradeoff between independence and interdependence. In our work, we draw mostly from the aspect of self-causality/identity\footnote{Hereafter, we refer to this aspect as simply ``self-causality.''}, which refers to the level and directness of a user’s decision-making and action execution in line with their own values. This concept is similar the concept of agency in cognitive science, which refers to the sense of deliberately controlling one's actions and affecting the world through those actions \cite{coyle2012agency}. From a more normative angle, prior literature has also argued that agency holds intrinsic value as a ``fundamental human need'' \cite{zhang2022screen}, a ``basic psychological need'' \cite{lukoff2021agency}, and a ``moral right'' \cite{reviglio2020thinking}. We integrate these perspectives into our work.

Reduction of user agency is a common concern on social media \cite{narayanan-algorithms, lukoff2021agency, harris-hijack}. A primary source of this concern is the opacity with which social media feed algorithms operate. Indeed, these algorithms are commonly referred to as ``black-boxes'' in HCI and social science literature \cite{christin2020ethnographer, pasquale2015black, bartley2021twitter, reviglio2020thinking, lustig2016algorithms}. While prior work in explainable AI for social media has attempted to open the black-box and make the algorithms more understandable to lay users \cite{ehsan2021xai, shang2022rec, kou2022explainable, kirchner2020countering, lai2022moderation, arnorsson2023reading}, even a fully transparent ``glass-box'' algorithm may still reduce agency. For one, attempts to explain a complex algorithm may trigger information overload, weakening users' decision-making abilities \cite{poursabzi2021interpret}. Transparency also does not guarantee self-causality---users may watch and understand the inner workings of the algorithm without any opportunity for control \cite{lipton2018mythos}. Transparency aside, widespread user behavior such as mindless scrolling \cite{purohit2023diets} and dissociation \cite{baughan2022dissociation} are telltale signs of users' agency loss when interacting with algorithmically-driven feeds. 

Dwindling agency has led to users deriving ``algorithmic folk theories'' to make sense of their social media experiences  \cite{eslami2015assumed, devito2018folk, devito2017twitter, karizat2021tiktok, siles2020folk, eslami2016folk}. Examples of folk theories include that users will see more content from friends who are more similar to them in their Facebook News Feed \cite{eslami2016folk} or that the TikTok For You Page algorithm prioritizes videos that feature aesthetics associated with wealthier lifestyles (e.g., large houses) \cite{karizat2021tiktok}. Theories may emerge from both \textit{endogenous} (originating within the platform, such as content patterns, friend count, likes, and comments \cite{bernstein2013invisible}) as well as \textit{exogenous} (originates outside the platform, such as user location \cite{burbach2018preferences}) information \cite{devito2018folk, rozenblit2002misunderstood}.

Upon formulating their theories, users applied them to better align
%(or ``domesticate'' \cite{simpson2022tame})
the algorithm to their preferences through strategic in-feed interactions. For example, Facebook users regularly visited profile pages of those whom they wanted to see more of and sought out more opportunities to tag them in posts, to signal their preferences to the algorithm  \cite{eslami2016folk}. To attempt to spread a video to other users' For You pages, TikTok users watched videos multiple times (even when they understood it perfectly the first time through), left longer comments, and hit the like/share buttons repeatedly even though they could only like or share a video once \cite{karizat2021tiktok}. On Twitter, users systematically created and shared content with certain words omitted or misspelled to evade the surveillance of the algorithm \cite{burrell2019twitter}. Users, however, remained uncertain of the efficacy of their techniques and even considered their actions to be manipulative and forced \cite{burrell2019twitter, eslami2016folk}. 

In our work, we seek to bring forth interactive tuning as a core component of social media algorithms, instead of treating such interactions as under-the-radar workarounds for preference and value alignment. We frame these interactions as teaching to highlight two essential qualities: naturalness and agency. As teaching is an inherent human ability that we both perform and receive throughout our lives, we are already naturally acquainted with its methods and mental models. Additionally, teaching is an agential activity---teachers are in charge of the teaching curriculum, delivering concepts to students, and evaluating student performance. Given this, how might we leverage our well-acquainted mental models of teaching in our online algorithmic interactions? We look to the paradigm of interactive machine teaching \cite{ramos2020interactive, ng2020knowledge, wall2019using} for inspiration. \looseness=-1

\subsection{Human Values in Recommender Systems}
Here, we focus on a particular class of content algorithms ubiquitous on social media and other everyday applications---recommender systems. A growing body of work proposes techniques and broader calls to action to embed human values into recommender systems \cite{Stray2022BuildingHV, Stray2020AligningAO, ovadya2023bridging, Dean2022PreferenceDU, Friedman2023LeveragingLL, Leqi2023AFT}, drawing upon the field of value-sensitive design \cite{friedman1996value, borning2012next}. Borning and Muller define ``value'' as ``what a person or group of people consider important in life'' \cite{borning2012next}. We borrow this definition of value in our work; we use the term ``values\footnote{Note the distinction between ``values'' and the ``value'' of a post. The latter is defined in Section \ref{s:intro}.}'' to abstractly refer to a unique set of important considerations of an individual or group, and ``preferences'' to refer to particular considerations within that set. 

Many modern recommender systems \textit{infer} human values by implicitly learning them from user information and interaction history \cite{Stray2022BuildingHV, narayanan-algorithms}. Commonly tracked attributes for implicitly eliciting values include tracking clicks \cite{zhao2019youtube}, content dwell time \cite{Yi2014BeyondCD}, and affinity with other users \cite{Kang2022FromWY, edgerank}. Implicit learning can be desirable as it reduces interface-level friction and allows for more effortless and rapid consumption of content \cite{Knijnenburg2011EachTH, Chen2021ValuesOU}. Additionally, advancements in deep reinforcement learning have improved the robustness and sophistication of implicit learning to the point that learned systems are, at times, capable of truly reflecting users' values \cite{Chen2021ValuesOU, Lawo2021BuyingT}. For example, users were startled by the accuracy with which TikTok's For You Page algorithm could capture their preferences without any explicit signals \cite{simpson2022tame}. However, an important downside of implicit learning is its inability to facilitate user agency.
%Accuracy, however, is far from being the primary concern when compared to the lack of user agency that arises as inevitable byproduct of implicit learning.
A 2021 investigation of TikTok's algorithm found that it quickly led users down niche content rabbit holes towards ``fringe'' content \cite{wsj-tiktok}. Others have raised concerns about recommender systems' ability to distort speech \cite{Stray2021WhatAY, saveski2021structure}, exacerbate polarization \cite{asimovic2021testing}, discriminate users and creators \cite{lambrecht2019algorithmic}, and erode mental health \cite{Stray2020AligningAO, lup2015instagram}. \looseness=-1

Given the concerns around implicit learning, many have
%derived methods and interfaces to empower users to instill values into
sought to develop recommender systems with \textit{explicit controls} \cite{Jannach2016UserCI, Knijnenburg2011EachTH, Ooge2023SteeringRA, Liang2023EnablingGE, Dooms2014ImprovingIM, Petrescu2021MultiStepCU, Friedman2023LeveragingLL, Jasim2023EditableUP}. Users have reported higher levels of satisfaction \cite{jin2017different, Kim2023InvestigatingHU}, trust \cite{Ooge2023SteeringRA}, and engagement \cite{Liang2023EnablingGE, Jannach2016UserCI} when they were given opportunities to exert control over the system, even when the controls had no impact on the output \cite{vaccaro2018illusion, bennett2023agency}. Examples of explicit controls include thumbs up/down buttons to rate recommended items \cite{zhao2013interactive, He2016InteractiveRS}, sliders and toggles for adjusting desired content characteristics \cite{jhaver2023personalizing}, drag-and-drop topic specifiers \cite{Dooms2014ImprovingIM}, keyword critique \cite{Petrescu2021MultiStepCU}, and manual selection of the recommender algorithm \cite{Ekstrand2015LettingUC, bluesky-feeds}. More recent works have leveraged the semantic capabilities of large language models to enable the expression of preferences conversationally through a chat interface \cite{Friedman2023LeveragingLL} and through editing a natural language user profile \cite{Jasim2023EditableUP}. Explicit controls come with their own set of challenges---users may not know that these controls exist or what they do \cite{hsu2020control, singh2020charting, jhaver2023personalizing}, find them cumbersome to use and keep up-to-date \cite{Jannach2016UserCI}, or do not see value in engaging with them \cite{Konig2022ChallengesIE}. Indeed, prior work showed that most users prefer a hybrid approach that combines implicit and explicit learning \cite{Knijnenburg2011EachTH, Mendez2017UserGA}. One promising approach in this vein is to design controls that allow for simultaneous expression of direct feedback and less direct social signals; real-world examples include ``react'' options on Facebook and LinkedIn \cite{Stray2022BuildingHV}. 

A common theme across both implicit and explicit learning is that particular content features are assumed to be more important to the user---features that the system then learns on. However, few works have questioned whether those features are truly ones a user cares about or wants the system to learn when they consume content. A news recommender may accurately capture (implicitly, explicitly, or some combination of both) a user's preferred article length, but will still fail to align with user values if the user does not care much for article length. In our work, we seek to elicit \textit{which content features users truly value} in social media feeds to orient future work in value-sensitive recommender systems.

\subsection{Interactive Machine Teaching}
Interactive machine teaching (IMT) is an interaction framework by which subject matter experts---who are often not experts in machine learning---draw upon their personal expertise to train machine learning (ML) models that can operate effectively within their domain \cite{ramos2020interactive, ng2020knowledge, jorke2023pearl, feng2023iml, zhou2022gesture, wall2019using}. While conventional ML is primarily concerned with developing algorithms that automatically learn conceptual representations from training data, IMT argues that learnable representations should directly come from human knowledge \cite{wall2019using, ng2020knowledge}. This way, users feel more agency while maintaining a firmer grasp of what the model learns, making models more transparent and debuggable \cite{ramos2020interactive, Kim2021TeachingLearningIA}. IMT consists of three main stages that form a ``teaching loop'' \cite{ramos2020interactive}:

\begin{enumerate}
    \item \textbf{Planning:} the human teacher (the subject matter expert) identifies a task for the algorithmic learner (the machine learning model) to complete, along with a curriculum (set of examples and representations to help teach the learner, typically in the form of a small dataset).
    \item \textbf{Explaining:} the teacher shows the learner examples and explicitly identifies concepts the agent should learn.
    \item \textbf{Reviewing:} the teacher allows the learner to predict some unseen examples, corrects any erroneous predictions, and updates the teaching strategy/curriculum accordingly.
\end{enumerate}

Central to IMT is the \textit{teaching language}, an interface by which the teacher crafts expressive representations that communicate desired concepts and are learnable by an algorithmic agent \cite{ramos2020interactive}. %For example, in Pearl \cite{jorke2023pearl}, an IMT tool that helps users reflect on personal calendar data for workplace wellbeing, possible teaching languages include labels that the user can apply to blocks of time on their calendar as well as user-defined rules (e.g., meeting activity during the day means working). 
In order to derive and use a teaching language, users must engage in a practice known as \textit{knowledge decomposition}. Ng et al. \cite{ng2020knowledge} define knowledge decomposition as ``a process of identifying and expressing useful knowledge by breaking it down into its constituent parts or relationships.''%For example, in a case study of workplace wellbeing, the user may have an overall goal of \textit{improving work-life balance}, which can be decomposed into a relationship between \textit{keyboard strokes after 5pm} and \textit{working after hours} \cite{jorke2023pearl}.

IMT shares similar goals with the paradigm of reinforcement learning
% (or its supervised learning sibling, active learning \cite{settles2009active}),
and, more specifically, reinforcement learning with human feedback (RLHF) \cite{ouyang2022rlhf}.
Both IMT and RLHF strive to interactively and iteratively embed specialized human knowledge into a machine-learned system. In RLHF, an algorithmic agent strives to find a \textit{policy} (a map of the agent's states to actions) that maximizes some long-run measure of reinforcement as defined by a \textit{reward function} in a dynamic environment \cite{sutton2018reinforcement, kaelbling1996reinforcement, christiano2017deep}, after which human feedback is be used by the agent to fine-tune the model to act in accordance with the user's intentions \cite{ouyang2022rlhf, stiennon2020learning}. %Indeed, both IMT and RLHF rise to the challenge of aligning ML systems with human values, a challenge dubbed as ``the Alignment Problem'' \cite{christian2020alignment} and one that has captured significant interest from various academic communities with the rise of large language models \cite{openai-alignment, superalignment}. 
The key difference between IMT and RLHF, however,
is that in IMT, the human teacher has agency---defining the curriculum, explaining concepts, and evaluating performance---whereas in RL(HF), the human is merely an oracle that the agent queries to guide its decisions \cite{christiano2017deep}.
% lies where agency is situated during the learning process. In IMT, the human teacher is in charge---defining the curriculum, explaining concepts, and evaluating performance---whereas in RL(HF), the agent determines an optimal policy while querying the human as an oracle to guide its decisions \cite{christiano2017deep}.
While we see potential in both IMT and RLHF to better align a feed curation agent with user values, we seek to center user agency throughout the feed curation experience. As such, we focus on IMT and defer exploration of RLHF approaches to future work.

How can IMT inform end-user empowerment in social media feed curation? Applications of IMT in social media contexts so far have been limited. Jahanbakhsh et al. \cite{jahanbakhsh2023exploring} employed an iterative loop bearing some resemblance to IMT to train a personalized AI capable of learning and predicting a user's misinformation assessments, but identifying misinformation is just one of many salient dimensions for users when consuming content in social media feeds. Additionally, higher-level tasks such as misinformation assessment may be composed of lower-level observations such as low-quality images or suspicious links that can be informative characteristics for curating content in other scenarios. Our paper seeks to draw out such observations through knowledge decomposition to articulate the design space for teachable social media feed experiences.

\section{Method}
In this work, we map out the design space for teachable social media feed experiences through the lens of IMT. Specifically, we sought to understand \textit{what} or \textit{how} users would teach a personalized feed curation agent. To answer our research questions, we first designed a study inspired by techniques in knowledge decomposition \cite{ng2020knowledge} to obtain taxonomies of salient signals for assessing the value\footnote{In this work, we define value broadly based on what participants want to see before others. Posts that participants want to see first first in their feeds has the highest value.} of the posts in a social media feed. We supplemented our taxonomies with qualitative interview data from the study. 

\subsection{Signal Elicitation Study}
\subsubsection{Participants}
Our study was conducted with 24 regular social media users between the ages of 18--65. We recruited some participants through the platforms of interest in the study (Instagram, Mastodon, TikTok, and Twitter; see Section \ref{s:platforms} for more details) and others through our institution's Slack channels and word-of-mouth. Prospective participants indicated which of the four platforms of interest they used, their frequency of use for each platform, and basic demographic information such as age range and gender, in our screening form. We started by accepting participants who used their platform(s) at least a few times a month, on a first come, first-serve basis. Selected participants then chose one of the four platforms of interest to use for the study. Later on, because we wanted similar numbers of participants for each platform, we only sent invites to participants who had a higher likelihood of choosing a platform for which we were seeking more participation, based on their indicated frequency of use for that platform. We stopped recruiting when we reached data saturation across all platforms. 

In the end, we had an even distribution of participants across our 4 platforms, such that there were 6 participants per platform. The far majority of participants (20) used their platform of choice daily, while 3 used it a few times a week and 1 used it a few times a month. Half of our participants (12) were aged 18--24, 6 were 25--34, 3 were 35--44, 3 were 45--54, and 1 was 55--64. 14 identified as women, 8 as men, and 2 identified as non-binary. All had at least 3 years of experience on social media, with all but one having at least 5 years and just over half (14) having at least 10 years.

We note that our participant pool may not be a representative sample of social media users, neither on a per-platform basis nor in aggregate. However,
% this is not of concern because
our study's intention is to map out a design space illustrating some interactive possibilities for teachable feed experiences, rather than to make generalizable assertions about
the relative frequency or importance of observed interactions.
% which interactions are more important and should therefore be implemented.
We also recognize that our resulting taxonomy is not exhaustive and new signals may emerge from specialized use cases and communities.

\subsubsection{Study setup and procedure}
\label{s:study-procedure}
\begin{figure}[h]
    \centering
    \includegraphics[width=0.5\textwidth]{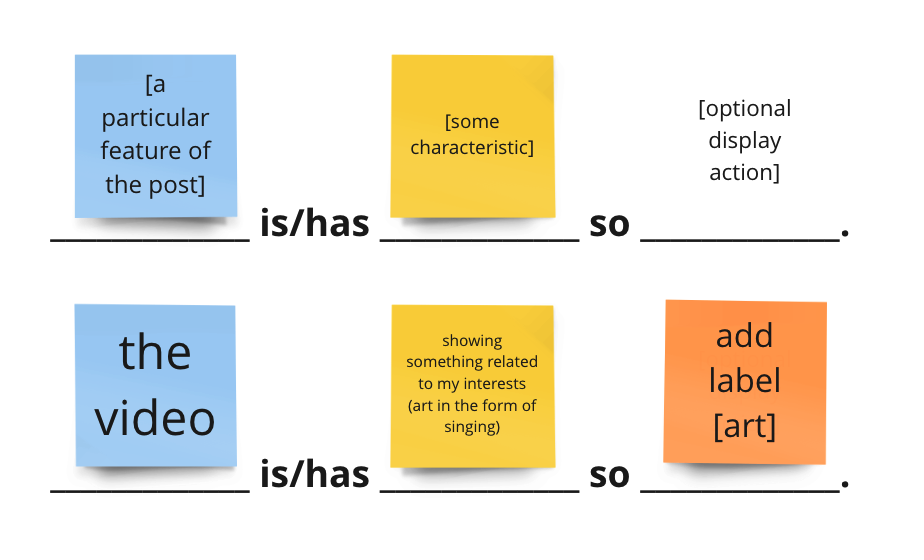}
    \caption{Examples of signals from our study. On top is a default signal template that we provide to users, and on the bottom is one completed by a participant.}
    \label{fig:signals}
    \Description{Two signals, one on top of the other. Each signal consists of a text element that reads: ___ is/has ___ so ___. The one on top contains a blue sticky note in the first blank with text that reads: [a particular feature of the post]. It also has a yellow sticky note in the second blank that reads: [some characteristic]. It contains plain text in the third blank that reads: [optional display action]. The signal on the bottom contains a blue sticky note in the first blank with text that reads: the video. It also has a yellow sticky note in the second blank that reads: shows something related to my interests (art in the form of singing). It contains an orange sticky note in the third blank that reads: add label [art].}
\end{figure}

To formalize the knowledge decomposition process \cite{ng2020knowledge}, we defined a \textit{signal} as an information unit consisting of two primary components: a \textbf{feature} and a \textbf{characteristic} (see Fig. \ref{fig:signals}). A \textbf{feature} is a class of information that can be extracted from a post. Example features include the post's author, the textual content, image(s), the number of likes, and the post's topic. A \textbf{characteristic} is a subjective statement that describes a feature. It is subjective in the sense that its significance may vary between participants. For example, the feature ``the post's author.'' is described by the characteristic ``is someone I know from in-person interactions.'' A third, optional component, an \textbf{action}, is something the participant can perform to the post in response to a feature-characteristic combination. For example, if the post author is someone the participant knows through in-person interactions, they may choose to take an action of ``trigger a notification'' for posts from that author. Actions were an optional component of a signal and could also be triggered by any combination of feature-characteristic pairs.

Participants were asked to compose signals (Fig. \ref{fig:signals}) and arrange them on a Miro board.\footnote{Miro is an interactive and collaborative virtual whiteboarding tool \cite{miro}.} Our study was conducted virtually, 1:1 through Miro and Zoom. We invited each participant onto a copy of the board and interactively co-authored signals with them in Miro while communicating through Zoom. Before the start of the study, participants submitted 10 screenshots of content on their own social media feeds from a platform of their choice out of Instagram, Mastodon, TikTok, or Twitter. We specifically asked for the first 10 posts they saw (and were comfortable sharing with us) in their ``home''\footnote{On Instagram and Mastodon, this refers to the main feed users see when they log in. For TikTok, we equated the home feed with the For You Page. For Twitter, we equated it with the Following feed, as Twitter was just starting to roll out its For You feed as we were conducting our study.} feed. Participants' posts were privately confined to their board and were not shared with anyone beyond the research team. 

\begin{figure*}[h]
    \centering
    \includegraphics[width=1\textwidth]{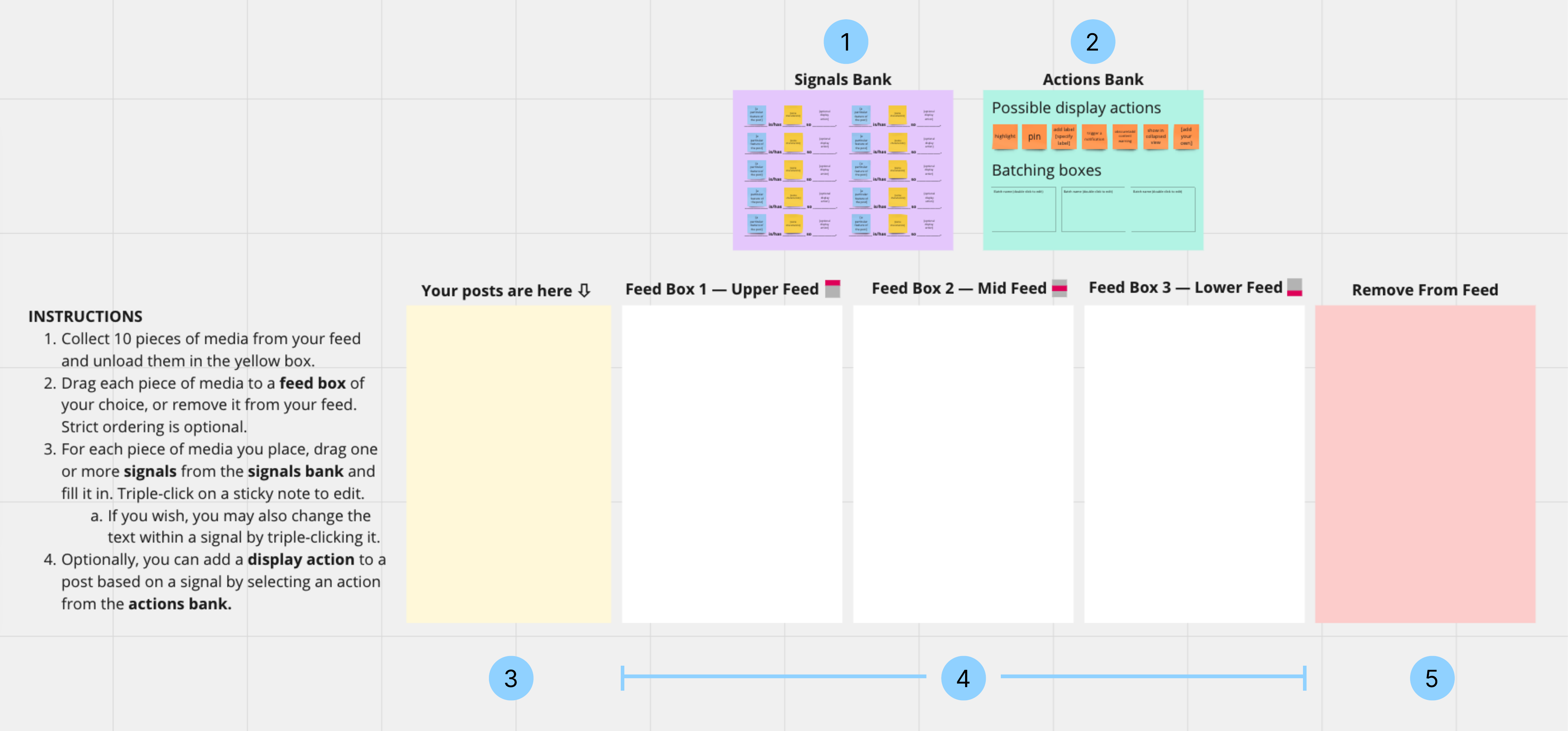}
    \caption{Areas of the Miro board. \textbf{(1):} signals bank with editable signal templates. \textbf{(2):} actions bank with sample actions and boxes for grouping content. \textbf{(3):} area where we uploaded participants' posts prior to the study. \textbf{(4):} the upper, middle, and lower feed boxes. \textbf{(5):} area for placing content that the participant would like to be removed from their feed.}
    \label{fig:board}
    \Description{A workspace on Miro, a collaborative whiteboarding platform, with a gray background and colorful rectangles (2 on top and 5 on the bottom) to indicate regions where activities will take place. One of the rectangles on top is called the Signals Bank and contains 10 signals while the other is called the Actions Bank and contains actions one can take written in text on top of orange sticky notes, along with boxes to batch together content. Instructions for the activity are written in text on the left side of the board.}
\end{figure*}

\begin{figure*}[h]
    \centering
    \includegraphics[width=1\textwidth]{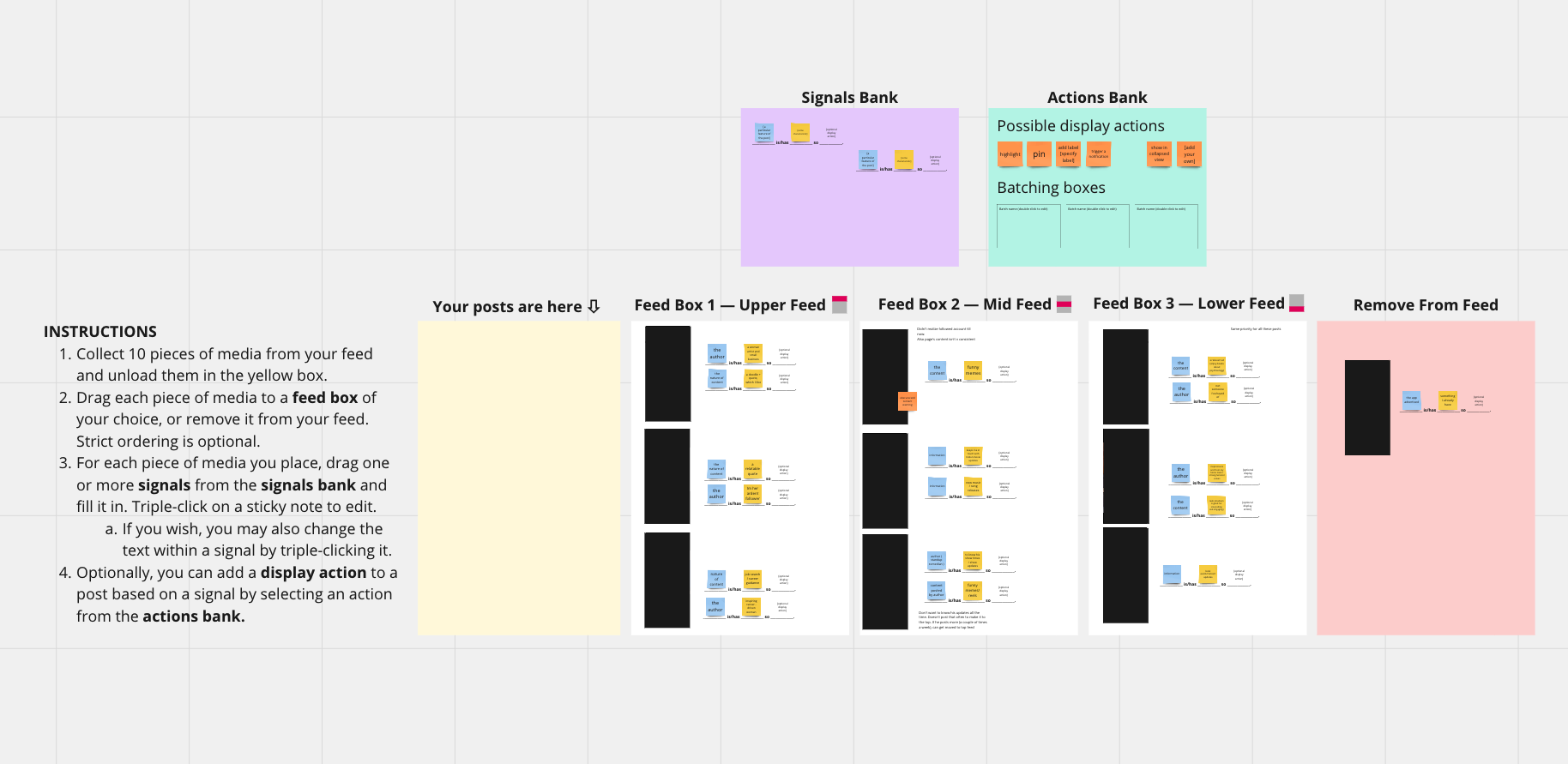}
    \caption{An example of a Miro board after a participant has completed the signal elicitation study. The posts themselves have been obscured by the research team to preserve the participant's privacy.}
    \label{fig:completed-board}
    \Description{A workspace on Miro, a collaborative whiteboarding platform, after the signal elicitation activity has been completed. The structure of the board is the same as the previous figure. There are 10 screenshots of social media posts (the contents of which have been redacted for privacy) spread out among 4 rectangles. Each screenshot has 1--2 completed signals next to it. }
\end{figure*}

The study board was separated into 5 main areas; Fig. \ref{fig:board} shows the board in its initial state. We populated the 10 screenshots participants submitted to us to area (3) in a random order before the start of the study. Area (1) contained editable signal templates while area (2) contained optional elements such as actions and boxes for participants to visually group posts they want to keep together. Area (4) was a space for participants to construct an ``ideal'' feed by organizing their posts into an upper, middle, and lower feed. While a linear feed may not have this exact distinction, we used it to roughly separate participants' perceived value of posts into 3 categories of descending value. %Participants dragged their content from area (2) to one of the 3 feed sections. 
Finally, area (5) was where participants could place any content that they would like to remove from the ``ideal'' feed they were assembling. An example of a Miro board after the activity has been completed can be seen in Fig. \ref{fig:completed-board}.

Using this board, we first asked participants to drag their content from area (3) into one of the 3 feed sections in areas (4) or area (5). We then asked them to further elaborate on their organization (why they value or do not value a particular post) by writing at least one signal using the templates in area (2) for all 10 posts. We co-authored these signals with participants by collaboratively editing the text in the signal template until participants were satisfied that the signal accurately represented their perspectives. Because the signals conformed to given templates, we relied on participants' think-aloud dialogue, which was recorded and transcribed, to capture their more nuanced reasoning. Many participants wrote more than one signal per post as there were multiple features they took into consideration. After all 10 posts were associated with at least one signal, we allowed participants to optionally assign actions to the signals. \looseness=-1

Upon conclusion of this interactive activity, we conducted a brief interview where participants reflected on their most frequently mentioned features and characteristics, as well as general experiences and desiderata with social media feeds. In total, the study took around 45 minutes to complete.

Our study was reviewed and approved by our institution's IRB under Study \#00016757. All participants received a \$20 USD gift card after completing the study. All studies were conducted virtually on Zoom between January and April 2023, audio recorded, and transcribed. 

\subsubsection{Platforms}
\label{s:platforms}
Our participants submitted content from their choice of one of 4 platforms: Instagram, Mastodon, TikTok, and Twitter. We were interested in these platforms as they were already popular or were gaining popularity at the time of the study, and also varied along two dimensions of interest: content format and engagement optimization. \textit{Content format} refers to the primary format with which content is displayed and consumed on the platform. For Mastodon, this was text, while Twitter uses a combination of text and visual media (images and videos). Instagram is mostly image-based, while TikTok is mostly video-based. \textit{Engagement optimization} refers to the extent to which a platform draws from its broader content pool based on a user’s prior engagement with a piece of content, rather than relying solely on those the user follows. Mastodon is strictly reverse chronological and has zero engagement optimization, while Twitter provides users with a choice of viewing an engagement-optimized ``For you'' feed and a more lightly optimized ``Following'' feed. Instagram offers an engagement-optimized home feed by default. TikTok is well-known for its engagement optimization on its For You feed \cite{wsj-tiktok}. 

We recognize that users may choose to engage with these platforms for different reasons---for example, one may log onto Instagram to share pictures from their personal life and catch up with friends, while only logging onto Mastodon for professional networking. We see this as a potentially rich source of insight in our study. Feed design and affordances are heavily influential in shaping perceptions of what a platform is best used for \cite{Kender2022TheSO}, and hearing participants' varied interaction strategies across different use cases can help us better envision the design principles and affordances users may seek given a particular use case or goal.

\subsection{Data Analysis}
Our study generated data in the form of 411 signals (with some repetition within and between participants) and 24 interview transcripts. We analyzed the two data sources separately while noting complementary and contrasting themes between the two.

\subsubsection{Signal data}
After all studies were completed, the first author aggregated signals across all participants and processed them into a spreadsheet to preserve pairing between features and characteristics. Within each group, the first author separated features or characteristics referring to an \textit{account} from which a post was made from those referring to a post's \textit{content},\footnote{This step was necessary for a coherent analysis as the two were fundamentally distinct (see Section \ref{s:taxonomies})} looking up the pairing in the spreadsheet as necessary. 

We used a hybrid approach of inductive and deductive coding  \cite{fereday2006demonstrating} to transform our data into taxonomies. The first author inductively sorted features and characteristics into broad themes through affinity diagramming; this process was independently performed on features and characteristics. Results from our inductive analysis were discussed and iterated upon with the broader research team at weekly meetings. For features specifically, the first author referenced common information available on interfaces of social media posts (e.g., username, handle, like/reshare/comment buttons) to further code the features in a deductive manner. Initially, the coding was not informed by the need to accommodate multiple platforms and some platform-specific features, such as lists on Twitter and Mastodon, were included. However, it became clear as coding proceeded that the vast majority of features were shared across all platforms in our study. As a result, we incorporated platform-specific codes into platform-agnostic ones---for example, the ``list'' code was split up and merged into multiple account-related features depending on how participants discussed their use of lists. We instead relied on our interview data to capture discussions of platform-specific experiences.

Our initial round of coding produced 7 and 16 account-related features and characteristics, respectively, and 20 and 31 content-related features and characteristics, respectively. The results seemed excessively granular for a taxonomy, so we iteratively grouped our codes into higher-level labels, discussing with the rest of the team as necessary.  Labels that fell outside of this paper's scope were also eliminated. Our final set of labels contain 4 and 10 account-related features and characteristics, respectively, and 9 and 18 content-related features and characteristics, respectively. Features and characteristics defined the vertical and horizontal axes of our taxonomies, respectively. We did not include an analysis of the ``action'' component of signals as participants used them sparingly in the study. The definitions for all labels in our taxonomies can be found in our Supplementary Materials. 

The intersection of each feature-characteristic theme was tallied and marked with either ``\textbf{$+$}'' to indicate that this combination had positive value (increased a post's perceived value),  ``\textbf{$-$}'' to indicate negative value (decreased perceived value), or ``\textbf{$\circ$}'' to indicate that there was no consensus among participants. Although we tallied exact counts, we expressed them as broader value ranges using saturation maps in the taxonomies to embrace the qualitative nature of our data and avoid invoking notions of generalization or numerical comparisons based on our counts.

% [Taxonomy distillation]

After our aggregate taxonomy with data from all platforms was finalized, the first three authors then completed the same taxonomy using disaggregated, platform-specific data. We separated the signals by platform and deductively coded the features and characteristics using the themes we developed from the aggregate data, resolving any disagreements in weekly data analysis meetings. % As these codes already existed from the aggregate analysis and this data was merely a subset of the aggregate data, no inter-coder reliability metric was needed. 

\subsubsection{Interview data}
\label{s:interview-analysis}
We took an inductive approach to performing thematic analysis on our interview data. Interview transcripts from the studies were initially auto-generated from Zoom recordings, after which the first author manually reviewed them while referencing the audio recordings to correct any incorrectly transcribed text. The first author took a first pass of open coding over the data, identifying insightful regions of the transcript and possible themes for further analysis. The first three authors then developed a codebook containing themes from open coded data that the team agreed were relevant and interesting. This codebook was used to collaboratively code the transcripts, with disagreements resolved through discussions among coders. Coders also wrote summary memos from the themes and discussed them with the broader team at weekly meetings, iterating on the themes and memos as necessary. Initially, the codebook consisted of 7 high-level themes synthesized from over 20 sub-themes. After some deliberation among the team, some high-level themes were combined and sub-themes were remixed to form the 5 high-level themes in the final codebook. Our final codebook is shown in Table \ref{t:themes}.  

\begin{table*}[h]
\centering
    \begin{tabular}{p{5cm} p{9cm}}
    \toprule 
    Theme & Description\\
    \midrule 

    Passive and active use of feeds & Participants' engagement in passive browsing, active consumption, or a combination of the two patterns.  \\

    People- vs. content-centric feeds & Participants' (often platform-specific) preferences for people-centric or content-centric feed experiences. \\

    Unactualized value of saved content & Participants' strategies and pain points in using existing archival features to to revisit content on platforms. \\

    Nuanced evaluation of posts & Participants' consideration of positive and negative facets of a post, as well as factors outside the post context, when judging a post's value. \\

    % Value effects of outside context & Outside context, such as off-platform content consumption or mood, had an impact on participants' content preferences. \\

    Self-causality and lack thereof & Participants' experiences with (the lack of) algorithmic control and transparency while curating content.  \\
    
    \bottomrule
    \end{tabular}
    \caption{Our 5 themes based on analysis of the signal elicitation study participants' responses  }
    \label{t:themes}
    \Description{A 2-column table with headings Theme and Description. There are 5 rows with the 5 high-level themes alongside their descriptions.}
\end{table*}

\section{Findings}
We begin this section by presenting our taxonomies for account- and content-based features and characteristics. We then showcase findings from our transcribed data, some of which complement our taxonomy data while others reveal what our taxonomies could not capture by themselves. Throughout this section, we use a two-letter abbreviation (IG for Instagram, MA for Mastodon, TT for TikTok, and TW for Twitter) to indicate a participant's chosen platform in the study.

\subsection{Taxonomies}
\label{s:taxonomies}

% \url{https://docs.google.com/spreadsheets/d/1Lnvo6D8ayZmdjyOUjnGth3W_QYfmhtbLo6c4VRc2jZU/edit?usp=sharing}

Our analysis yielded two distinct sets of taxonomies: account-based (with features and characteristics relating to accounts that post and engage with content) and content-based (with features and characteristics relating to content that accounts post). We made this distinction as the two sets were fundamentally different in nature---characteristics from one were incompatible with features from the other, and vice versa. Definitions for all labels used in our taxonomies can be found in our Supplementary Materials.

\begin{figure*}[h]
    \centering
    \includegraphics[width=0.65\textwidth]{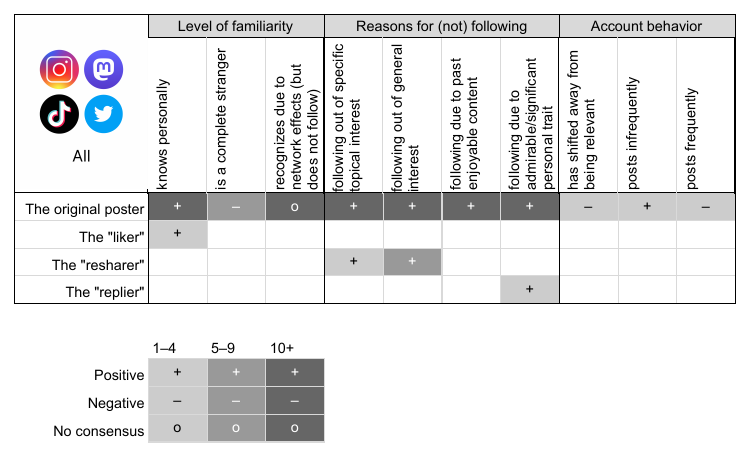}
    \caption{Our account-based taxonomy, aggregated across Instagram, Mastodon, TikTok, and Twitter.}
    \label{f:acct-taxonomy-agg}
    \Description{A gridded grayscale table with 4 rows and 10 columns, excluding row and column labels. The cells are either not filled in, or filled in with either light, medium, or dark gray. The cells that are filled in either have a $+$, $-$, or $\circ$ in them. The top row is filled in with mostly dark gray cells that contain $+$.}
\end{figure*}

\begin{figure*}[h]
    \centering
    \includegraphics[width=1\textwidth]{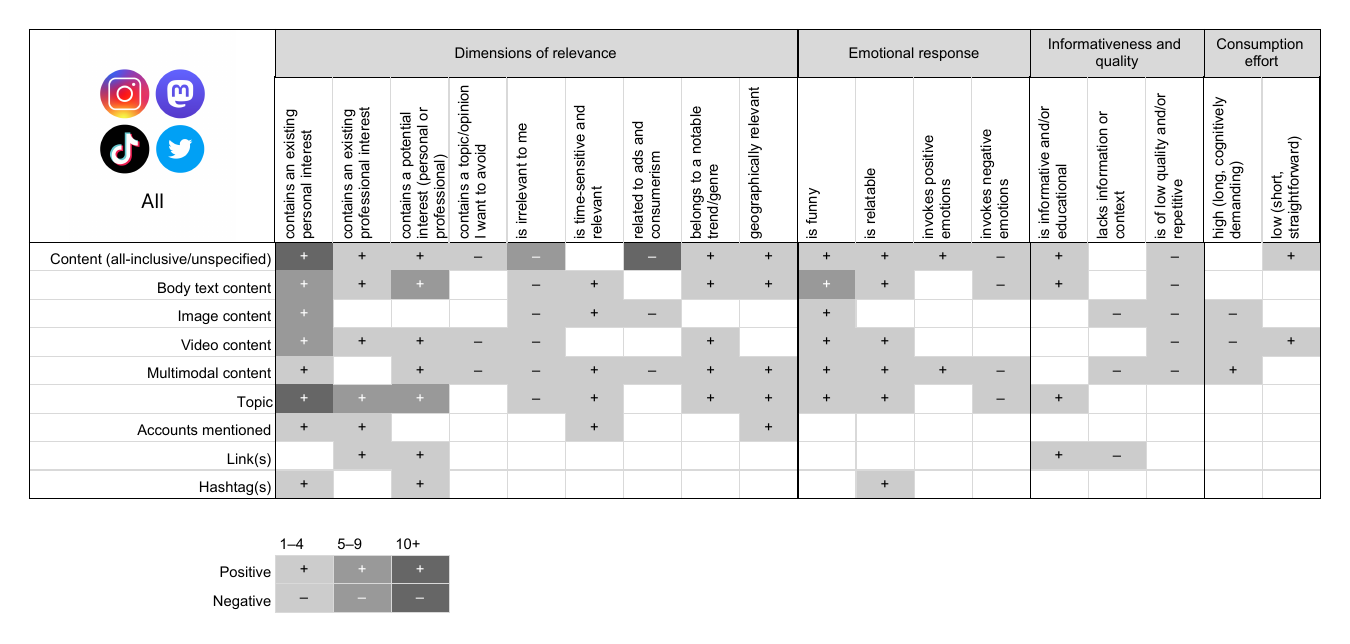}
    \caption{Our content-based taxonomy, aggregated across Instagram, Mastodon, TikTok, and Twitter.}
    \label{f:content-taxonomy-agg}
    \Description{A gridded grayscale table with 9 rows and 18 columns, excluding row and column labels. The cells are either not filled in, or filled in with either light, medium, or dark gray. The cells that are filled in either have a $+$, $-$, or $\circ$ in them. Cells that are filled in are distributed throughout the table, mostly shaded light gray containing a $+$.}
\end{figure*}

Fig \ref{f:acct-taxonomy-agg} shows our account-based taxonomy with data aggregated across all four platforms. We see participants rely heavily upon the \textsc{original poster} as a feature when evaluating a post. This feature is described with a wide range of characteristics---that is, there are diverse reasons why participants may pay attention to the original poster. Some characteristics, such as \textit{``knows personally''} depend on personal context and even offline interactions the user may have had with the account holder. Other prominent characteristics were concerned with users' evaluation of the account holder's personal traits (e.g., if the account holder had a trait that the participant admired) or some trend in the content they released (e.g., alignment with the participants' topical interests). We note that recognizing an account from network effects is the only characteristic that garnered mixed valuations, dependent on whether the participant's network portrayed the account in a positive light. In our disaggregated, per-platform view of the same taxonomy (Supplementary Materials), we observe that TikTok's taxonomy is noticeably sparser than the others, indicating there is less interest in account-related features, a possible result of platform design. We dive into a deeper investigation of this with our transcript data (see Section \ref{s:findings-ppl-content}).
% define multimodal?

Our aggregate content-based taxonomy (Fig. \ref{f:content-taxonomy-agg}) shows that participants' evaluation of content is distributed over a much greater range of features when compared to our account-based taxonomy. Unsurprisingly, the most prominent characteristic associated with a negative evaluation is \textit{``related to ads and consumerism.''} \textsc{Topic} was a prominent feature contributing to positive evaluation, especially when the topic aligned with users' existing personal interests. However, topic differs from many of the other features as it is often \textit{interpreted} by the user rather than \textit{explicitly defined} in a post. Participants were also generally more drawn towards personal interests rather than professional ones in their feeds (except for Mastodon), although both were considered across all platforms. \textit{``High consumption effort''} is a noteworthy characteristic due to the mixture of positive and negative evaluations in the same column. While some participants enjoyed shorter content due to their ease of consumption, P4[TT] saw longer-form content as more likely to contain material of interest. In our disaggregated taxonomies, we observe that TikTok was the sole contributor of the characteristic \textit{``belongs to a notable trend/genre.''} This leads us to believe that participants are more attuned to the trendiness of content on TikTok than on other platforms.

Ultimately, our taxonomies show that participants' evaluations of posts from their feed were diverse and multi-faceted. Such preferences are unlikely to be captured in their full richness with a simple ``like'' or even a ``react'' on today's platforms, nor a one-size-fits-all approach to feed curation. We thus see potential in leveraging techniques from IMT to design new affordances that allow users to teach more nuanced preferences to the algorithm for more agential and expressive curation.

\subsection{Interview Findings}
In addition to writing signals, we asked participants to think aloud throughout the activity and reflect on their broader experiences with social media feeds in a post-activity interview. Here, we discuss salient themes that emerged from our transcript data. 

\subsubsection{Participants engaged in both undirected and directed use of their feeds}
\label{s:findings-passive-active}
Many participants described using their feeds to consume information in an undirected manner. We characterize \textit{undirected consumption} here as consumption without a specific goal in mind. Common scenarios in which participants engaged in undirected consumption include: (1) browsing the feed to be entertained (2) casually catching up on interesting conversations, and (3) instinctively opening the feed out of habit. %Many TikTok participants mentioned entertainment as the primary reason why they use the platform, in line with prior work \cite{Barta2021ConstructingAO, klug2021tiktok}. To P10[TT], TikTok was  \textit{``pure entertainment, almost like watching TV for me.''} 
This entertainment-induced passive consumption, which was especially common among our TikTok participants, can often lead participants to perceive content as more disposable. For example, P11[TT] noted that \textit{``if [a post] doesn't raise a question or doesn't seem like something I'm interested in, then I'll just scroll away because there's an infinite number of videos I can watch.''} Besides entertainment, participants also engaged in undirected consumption when casually browsing for interesting updates and catching up with relevant conversations. P2[TW] likened this to \textit{``hallway chats [...] or water cooler conversations.''} Finally, some admitted that their undirected browsing tendencies were simply driven by habit and instinct. P7[IG] shared that \textit{``if I'm waiting for something, my hand just automatically goes [to my phone].''} P23[TT] felt similarly for TikTok: \textit{``I feel like it’s more so a habit of [...] opening the app, and just hoping I find something interesting.''}

On the other hand, participants also mentioned many instances of using their feeds in a directed manner. We characterize \textit{directed consumption} as information seeking with a specific goal. Active consumption was more frequently linked to the activity of \textit{searching} rather than \textit{browsing}. P4[TT] discussed their use of TikTok as \textit{``kind of like a Google for me. I want to learn about, for example, these new car features, or `how do I bake a cake?' And what are some good restaurants in [my area]?''} %P8[IG] also talked about their use of TikTok to learn about meditation techniques: \textit{``I [searched] meditation techniques, and there was so many content on that and it was very easy to filter and sort information.''} 
P24[MA] even kept a dedicated search column in their multi-column Mastodon feed to search for particular hashtags. 

Given these characterizations, participants did not limit themselves to solely undirected or directed consumption. They would often use their feeds for both, switching between the two based on their goals and context. P15[TW] summarized this aptly:

\begin{quote}
    \textit{``If I am in a low focus parsing mode, then I will read everything in the timeline. If I want to focus
    and get things that are a little bit more important [...] or I’m trying to answer a question, I’ll say, `I care about this topic, or I care about [hearing from] these experts.' Let me dig in and see what’s going on here.'' (P15[TW])}
\end{quote}

P13[TT] mentioned that they will engage in undirected browsing \textit{``if I have time to fill''} but otherwise will actively seek or follow up on information, such as restaurant recommendations. P6[TW] used Twitter \textit{``very intentionally for professional [networking]''} but is otherwise just \textit{``scrolling and looking around.''} %Passive and active consumption were also by no means mutually exclusive---aspects of each were sometimes combined. We consider this \textit{active browsing} behavior to be a combination of active search and passive browsing. P24M described an example of this:

% \begin{quote}
%     \textit{``Is there a conversation going on where I need to reply to something? So that's usually like, okay, I got this time set aside [for that], and I know there was some notification. So let me go follow up on those.'' (P24M)}
% \end{quote}

% In a similar vein, P1M explained that they make an active effort to browse content from \textit{``people who are not like me, regardless of the content [they post].''} 

Our observation of participants' consumption behaviors not only highlights the need for feeds to accommodate diverse modes of browsing, but also enable users to fluidly move between them.

\subsubsection{Tension existed between desires for people- and content-centric feed experiences} \label{s:findings-ppl-content}

In recent years, social media platforms have shifted towards content-based recommendations in an attempt to more effectively drive user engagement, particularly for younger demographics \cite{fb-aging}. While it is not clear from our taxonomies (Figures \ref{f:acct-taxonomy-agg} and \ref{f:content-taxonomy-agg}) alone how participants prioritized account-based or content-based signals, our interviews shed more light on this. Many expressed a desire for more \textit{people-centric} experiences in their feeds---curating and viewing content based on individuals they care about rather than the contents of the post itself. 
To P17[MA], the post author can serve as a reliable indicator of relevance: \textit{``There are some people that I know 95\% of the things they say are relevant to what I'm doing.''} Indeed, when asked about the features they prioritized for post evaluation, all but 7 participants indicated an explicit preference for account-based features over content-based ones. \looseness=-1

% This people-centric mental model thus came into tension with platforms' increasingly content-centric recommendations. The most striking example of this tension was Instagram. Of the 6 Instagram users in our study, 4 saw no content from close friends or acquaintances in their first 10 posts of their home feed. P9I only realized this behavior during the study and was repelled by it: 

% \begin{quote}
%     \textit{``I don't have any posts here from---I just noticed---my friends, which used to be a thing I really liked about social media. Just finding out what my friends are up to and feeling connected. Wow! What a horrible development. How have I only just noticed?''} (P9I)
% \end{quote}

% P9I's quote illustrates the ``frogboiling'' effect that users may be subject to on platforms: by rolling out changes gradually and tactically, platforms can shift user behavior to align more with platform goals without provoking user protest. 
This people-centric mental model thus came into tension with platforms' increasingly content-centric recommendations. The most striking example of this tension was Instagram. Of the 6 Instagram users in our study, 4 saw no content from close friends or acquaintances in their first 10 posts of their home feed. P7[IG] felt that their feed hindered their ability to stay updated with those they cared about: \textit{``None of these 10 posts were posts about people I know [...] And I'd like to know what's going on in their lives.''} P12[IG] echoed this and the desire for prioritizing posts from specific people: \textit{``I would love if I could see what my friends post at the top.''} P8[IG] leveraged the `Close Friends' feature \cite{insta-close-friends} to combat the algorithm's content-based recommendations so that they could still \textit{``be connected to [my friends].''} 

Interestingly, 5 of the 7 users who did not express a preference for account-related features used TikTok for this study. For TikTok users such as P10[TT], their home feed (For You Page) is mostly filled with content from \textit{``people I don't follow, or I've never heard of.''} Unlike P17[MA], who trusted certain authors to post relevant content, P23[TT] was the opposite: \textit{``even if it wasn't [this creator], I would stay tuned to the specific topic she's talking about.''} %P13T even considered topic relevancy to be the most important feature to consider when evaluating posts. 
A couple reasons (or combinations thereof) may prompt deeper engagement with content on TikTok. First, the primary use of TikTok was for entertainment (see Section \ref{s:findings-passive-active}). As is the case with YouTube, TikTok users' focus on entertainment may lead them to be more concerned with content consumption rather than fostering social interactions with people \cite{khan2017social}. Second, the design of the TikTok interface may serve to prioritize content, more so than other platforms. As prior design analyses of social media UIs \cite{Kender2022TheSO} point out, TikTok offers an immersive content-viewing interface with minimal whitespace---unlike other platforms, all other features within a post are overlaid on top of a TikTok video.

Our takeaway here is that both people- and content-centric feed experiences can be desirable. On some platforms, users prefer content-centric experiences; on others, they prefer people-centric ones. The tension between the two comes into play when, on a particular platform, users desire one but their feed forces them towards the other. Thus, when formulating designs for teachable feed experiences, we aim to equip users with necessary tools to agentially express their desired type of experience, and tweak that experience as they see fit.

\subsubsection{Perceived value of saved content was often not actualized} \label{s:archive-retrieval}
Many participants used archival features (e.g., ``save'', ``bookmark'') on their platforms for a variety of reasons. For some, the times when they checked social media were not when they wanted to take action on a piece of content. For example, P2[TW] shared that they \textit{``usually check social media in the mornings and then I have to work,''} so they saved posts related to science fiction (a topic of their interest) to read after work. P7[IG] saved encouraging quotes for \textit{``when I'm sad, or when I really need a pep talk.''} Over several months, P7[IG] also saved career-related posts in preparation for the upcoming recruiting season. %P11T enjoyed saving content that they can personally try in the future, such as recipes and fashion trends: \textit{``If it's a recipe and  want to try it, then I'll probably save it to a folder, or if it's fashion and I want to try that the next time I'm putting together an outfit, I would save that.''}

It is important to note the distinction between simply \textit{allowing a user to save content} and allowing them to \textit{curate their content archive} such that they can easily retrieve desired content. While all platforms in our study offered the ability to save content, there is much less support for managing saved content so participants could actualize their value. To circumvent the issue of not being able to retrieve saved content, P8[IG] said they would sometimes send content to another person instead of saving it on the platform so that they could rely on that person's memory to help retrieve the content later on: \textit{``I would send to my husband a lot of the things I like [...] That's also a strategy for `watch later.'''}. It also does not help that saved content often has low discoverability in platforms' interfaces. As of the time of writing, saved content in Instagram is 4 taps away from the home feed and requires traversing a user's profile menu, an area reserved for options such as privacy settings and payment information.

In many ways, an archive of saved content resembles a feed, but is free of algorithms---all curation happens manually. P24[MA] considers the lack of algorithms to be a double-edged sword: \textit{``inside of [archives], I don't get the good of the algorithm, but at least I wasn't getting the bad stuff either.''} %P19M was enthusiastic about organizing their saved content \textit{``it's like a separate news feed, and if an algorithm could help and start tagging [the content], that would be amazing.''} 
While a sophisticated curation algorithm may be unnecessary, we recognize that an organized content archive can act as a collection of examples representative of user preferences, acting precisely like a curriculum in IMT. We further explore this idea in Section \ref{s:multi-feed}.

\subsubsection{Participants' evaluations of posts were nuanced}
\label{s:nuanced-evaluation}

Many current social media platforms have explicit and implicit means by which users can provide feedback to content algorithms \cite{Stray2022BuildingHV, narayanan-algorithms}. However, this feedback can fail to capture users’ consideration for complex factors that influence their overall assessment. Participants often gave nuanced evaluations of posts which took into account multiple feature-characteristic pairs contributing to both positive and negative evaluations of a post, or factors outside the context of the post.

More complex evaluations were frequently due to conflicting assessments of different aspects of the content, such as a post that is about an undesirable topic but provides value to the participant in another way. For example, P10[TT] said that they are not interested in the Grammy Awards, but still wanted to keep related posts in their feed \textit{``just to stay updated on it.''} Similarly, P16[MA] described how an undesirable post could still enrich their social media experience: \textit{``I wish to some degree that it wasn’t posted but at the same time, it does have value to me. It gives me information about the person and it makes it feel like a real conversation.'' }

Sometimes, users may give mixed assessments of the post content and the post author. A negative assessment of the content of a post may outweigh a positive assessment of the author (or vice versa) in evaluating the desirability of a post in their feed. For example, in response to an offensive Instagram post, P22[IG] noted that \textit{``there’s no way for me to tell Instagram, `don't show me pictures like this anymore,'\thinspace''} although they still wanted to see posts from the author and \textit{``don't really want to mute the person.''}
A negative evaluation of the content of a post may even impact a positive/neutral evaluation of the author, or vice versa. P1[MA] explained how the distasteful content of a toot on Mastodon led them to change their evaluation of the author: \textit{I might even unfollow this person for tooting this thing out [...] [the post was] in some ways useful in that, like `wow, I can't believe the person is buying into this garbage, I don't know if I can trust other stuff that they put in my feed.'} This suggests that our two taxonomies, which capture both account- and content-based evaluations of posts, respectively, are not strictly independent and should be considered in concert.

In other cases, however, a participant’s evaluation of a post was influenced by context not local to the post itself. For example, mood can affect a participant's social media preferences. P8[IG] preferred ``short and sweet'' posts when stressed, to avoid spending too much time on their phone, and P12[IG] only wanted to see \textit{``corny Instagram captions''} when they were in a good mood.
Such temporary preference shifts might extend beyond a user’s mood on a given day and instead apply across a larger span of time. P17[MA] noted that while they typically did not mind seeing professional content on their Mastodon feed, they had \textit{``been kind of overwhelmed with mid and senior career people recently.''} Echoing this need, P19[MA] suggested a potential feed customization feature where users \textit{``could just mute a category for a day, or for some period of time.''}

Users' evaluations of posts involve multi-dimensional and often conflicting assessments of features within a post, as well as outside factors, including mood. We see a need for teachable experiences that accept both structured (e.g., via UI buttons) and unstructured feedback (e.g., via natural language) over longer timescales. This way, users have more flexibility to indicate temporary or context-specific preferences for feed curation.

\subsubsection{Participants experienced a lack of self-causality in feed curation}
\label{s:algorithmic-transparency}

Many participants did not trust algorithms' ability to curate feeds per their preferences. While describing why they were not interested in a post from an author they usually appreciate, P2[TW] said that they \textit{``don't trust the algorithm to know''} that distinction. P9[IG] explicitly said that they \textit{``don't want algorithms to try to sort the best content for me''} and would rather do it themselves. 

P9[IG]'s statement demonstrated a desire for \textit{self-causality}, a key aspect of agency defined by the ability to make personal choices and see these decisions reflected in an outcome in line with their values and goals \cite{bennett2023agency}. Participants even prioritized self-causality above content enjoyment: P8[IG] said that while they sometimes enjoyed the algorithm's suggested posts, they \textit{``would put it towards the end [of the feed], because it's not something I've decided I want to see.''} 
This desire for greater control over feed content also led participants to take actions to experience self-causality, for example, by forming rudimentary feed training behaviors. P10[TT] shared that they \textit{``will like videos that I liked and dislike on the ones I don't want to see anymore''}, because they are \textit{``pretty sure the algorithm will learn what you like''.} 

Participants' efforts to experience self-causality were hampered by an inability to determine why the algorithm fed them specific content. Participants often evaluated this based on different dimensions of relevance of the content in question, as represented in our content-based taxonomy. When describing how platforms know what kinds of feeds users are interested in, P23[TT] said that they \textit{``think [the platforms] kind of guess''}, as they had a STEM-related feed that they said was \textit{``cool, but I don't know why [the platform] would pick that feed specifically for me.''} P10[TT] described how they were \textit{``somehow on autism TikTok, even though I'm not diagnosed with it or anything,''} which \textit{``feels weird''} as they \textit{``don't know how I got there.''} This uncertainty made it difficult for P10[TT] to use like/dislike signals to curate their feed, as they described the process as \textit{``not foolproof... I will still get fully random [posts].''} Although participants could see that the algorithm was not accurately learning their preferences, they did not know why it was learning in this way or how to teach it more effectively, limiting their self-causality.

While participants expressed some negative sentiment towards algorithmic feed curation, we see that this sentiment may not be caused by the presence of algorithms themselves, but rather the lack of self-causality experienced in current platforms. IMT was motivated by similar concerns, and has the potential to increase algorithmic teachability and user agency if applied to feed settings. 

We now proceed by synthesizing our study's findings through the lens of prior literature on IMT to inform design principles for teachable feed experiences.

\section{Design Principles for Teachable Feed Experiences}
A primary objective of IMT is to leverage an end user's subject-matter expertise to train a learnable agent that can effectively aid the user in fulfilling their desired goals \cite{ramos2020interactive, ng2020knowledge, zhou2022gesture}. However, teaching languages\footnote{Recall that a \textit{teaching language} is an interface by which the human teacher can expressively communicate desired concepts for an algorithmic agent to learn.} proposed in prior work, such as PICL \cite{ramos2020interactive} and Pearl \cite{jorke2023pearl}, may not translate smoothly to social media settings---they demand high-effort, meticulous interactions from users to provide meaningful labels and descriptions from data examples. Given that social media is designed to support fast-paced consumption of information, actively introducing excessive friction via conventional IMT interfaces may worsen the user experience rather than improve it. \looseness=-1

Based on our taxonomy and participant interviews, along with prior literature in IMT, we outline five design principles to set the stage for teachable social media feed experiences. We do so to extend the core principles of IMT to the modern social media landscape. In some cases, this extension directly builds off of classic IMT principles; in others, those principles are re-examined and reconfigured. Our principles may be used independently or (ideally) in combination.

\textbf{D1: Situate the teaching language within the feed.}
In IMT, teaching has conventionally been performed in isolation from the environment in which the teaching materials originated. For example, when creating a recipe classifier in PICL \cite{ramos2020interactive}, the user first imports a set of documents containing recipes before they can start teaching. Note that PICL is a separate environment from where users may encounter recipes in-the-wild, such as on websites.
% Similarly, when assembling a reflective journal using workplace calendar data in Pearl \cite{jorke2023pearl}, the system imports calendar data into a dedicated interface for teaching, rather than enabling teaching activities on top of the user's default calendar application itself.
A separate teaching environment can enable more feature-rich and expressive teaching languages, but it can also disengage users from the act of teaching in social media settings. For one, the additional effort required to move posts and organize them outside of the feed is already seen as burdensome by our participants (see Section \ref{s:archive-retrieval}). Additionally, it may not be in platforms' best competitive interest to support exporting a post and any informative metadata off the platform. We therefore situate the teaching language \textit{within the feed itself}. This way, we can directly leverage platforms' existing representations and users' existing mental models, while lowering the effort required to perform teaching. A key question then arises: what exactly do users' mental models of current platform representations look like in the context of feed curation? Our taxonomies shed valuable light on this; we can leverage salient features identified in our taxonomies to inform the design of our in-feed teaching language. \looseness=-1

\textbf{D2: Be available, but not intrusive.}
We heard from many participants that social media can act as a much-needed break or distraction in the middle of the day, a means of relaxation, and a casual time-killer when small pockets of idle time arise. That is, participants were satisfied simply by letting the algorithm entertain them. In these scenarios (described by P15[TW] as \textit{``low focus parsing mode''}), persistently demanding extra attention and effort via IMT may in fact worsen the user experience and force users to expend more energy than they desire. We saw in practice that many participants switched between this low-focus mode and a more attentive, information-seeking mode (see Section \ref{s:findings-passive-active}), and that this switching was largely dependent on difficult-to-predict factors such as mood and context. In light of this, we ensure that our teaching language is available, but unobtrusive. That is, the interface is easily accessible to users across the entire feed experience, but can also be easily dismissed or hidden when not needed. 

\textbf{D3: Embrace a multiplicity of feeds.}
Some of our participants questioned why so many feed experiences had to be constrained to a singular feed and expressed a desire for multiple feeds to better organize their content. P12[IG] touched on the oddness of having diverse content mix in their feed: \textit{``it's weird to go past someone's bikini photo and then, `5 people killed at a refugee camp.'''} Indeed, the range of characteristics used to describe features of posts as captured in our taxonomy corroborates this diversity. As a less extreme example, it may be jarring for users to see content that ``contains an existing professional interest'' if they expect to scroll through posts whose content ``is funny.'' P16[MA] mentioned that they would like to create different feeds for the various Mastodon instances they were on. P8[IG] likened their feed to a newspaper and suggested different sections: \textit{``news, updates, events, pop culture, etc.''} In fact, platforms have also started to explore multi-feed experiences and broadening of algorithmic choice. Many, including Twitter, now offer an engagement-based ``For You'' feed alongside a ``Following'' feed featuring content from followed accounts. One smaller social platform called Bluesky has introduced Custom Feeds, where developers can create feed algorithms to which users can browse and subscribe \cite{bluesky-feeds}. TikTok has also introduced topic feeds based on inferred user interests \cite{tiktok-feeds}. We consider this shift towards feed multiplicity a promising direction, especially when providing users with a means of organizing their IMT curriculum.

\textbf{D4: Seek structured and unstructured feedback.}
% structured vs unstructured feedback
IMT workflows have traditionally been scaffolded with affordances that elicit structured feedback (e.g., data highlighting and labelling). Our findings in Section \ref{s:nuanced-evaluation} and the range of both positive and negative feature-characteristic pairs in the taxonomy suggest the need for teachable algorithms that accept both structured and unstructured feedback in order to capture users' nuanced evaluation of social media posts, which corroborates existing research in IMT. For example, Jörke et al. \cite{jorke2023pearl} provided a form-like interface for specifying key teaching concepts in which some fields allow the selection of existing data tags while others allow the user to write freeform natural language. Likewise, Zhou et al. \cite{zhou2022gesture} noted the importance of relying on users' unstructured object show-and-tell gestures in addition to a structured UI.
This combination of structured and unstructured feedback is desirable in real-world applications as users may make nuanced judgements that cannot be captured with structured feedback alone, but also require some guidance to express concepts in a format parsable by an algorithmic learner. Parsing
%and deriving meaning from
unstructured input, however, is now significantly less challenging due to technological advancements in large language models. We take this into consideration when balancing structured and unstructured feedback. \looseness=-1

\textbf{D5: Enable teaching and evaluation at varying timescales.}
% refinement over time - user quotes, transparency, evaluation, reflection
Bennett et al. \cite{bennett2023agency} identified timescales of interaction---ranging from \textit{micro-interactions} (a few seconds or less) to \textit{episodes} (seconds to hours) to \textit{life} (days to years)---to be a key aspect of human agency and autonomy. When asked to articulate their preferences with limited access to their feeds during the study, many participants had difficulty doing so because their preferences \textit{evolved over time} based on the content they saw. As such, they expressed a desire to agentially refine their feeds' behavior over an extended time period. This also happened to be less cognitively demanding, as P6[TW] pointed out: \textit{``I don't have to intentionally be like, okay, I'm gonna sit down and coordinate everything.''}
Our participants also described how their preferences shifted temporarily based on factors like mood (see Section \ref{s:nuanced-evaluation}). 
Indeed, we can see the pitfalls of assuming static preferences and designing only for interactions at short timescales via the inefficacy of set-and-forget personal content moderation tools \cite{jhaver2023personalizing}. That said, classic debates in HCI over direct manipulation interfaces (UIs providing immediate control feedback through elements such as buttons and sliders) versus interface agents (systems that perform actions on behalf of users, often after learning user preferences over time) reveal that aspects of both are vital in information-dense environments \cite{shneiderman1997direct}.We thus explore designs that enable teaching and evaluation at varying timescales. Specifically, we aim to leverage teaching interactions afforded through direct manipulation as a familiar interaction pattern to ground evaluations of algorithm performance over longer timescales. In doing so, we situate the teaching language as not only a tool for the teacher to articulate key concepts, but also one that aids evaluation of the algorithmic learner.

\label{s:principles}

\section{Proposed Feed Designs}
We now propose three feed designs for teachable social media feed experiences that embody our findings and design principles. Our goal is not to constrain the design space of teachable feeds with these designs, but rather present them as sensitizing concepts---emergent ideas that help direct attention to promising topics or phenomena \cite{blumer2017wrong, zimmerman2010analysis}---to illuminate salient paths for future research. 

Note that the mockups illustrating our feed designs feature a generic microblogging platform, similar to Twitter or Mastodon. Our reasons for making this decision were twofold. First, we wanted to show how our designs can operate on platforms with a diverse range of feature and characteristic preferences, as opposed to ones like TikTok where a particular media modality (and therefore certain sets of features and characteristics) is significantly prioritized over others. Second, we tap into the increasing design and development interest in microblogging alternatives since Twitter's change in ownership \cite{techcrunch-mastodon, usatoday-mastodon}. Many platforms in this space, including Mastodon and Bluesky, are also open source, making experimentation with novel ideas more accessible than on closed platforms. 

\subsection{Exploded UI Views}
\label{s:exploded-ui}

In 3D diagramming, an \textit{exploded view} is one where the individual components of an object are shown slightly separated from each other as if a small explosion occurred at the center of the object. Exploded diagrams depict inter-component relationships and are thus commonly found in instructional manuals. We employ a similar concept in a post's in-feed UI to serve as a teaching language for the elicitation of more granular preferences. On current platforms, ``liking'' a post can signal that the user enjoys \textit{some feature(s)} in the post, but it does not clearly communicate \textit{what specific feature(s)} they found like-worthy. An exploded UI view aims to recover this information by allowing the user to specify features within the post that they find (un)appealing.

\begin{figure*}[h]
    \centering
    \includegraphics[width=0.8\textwidth]{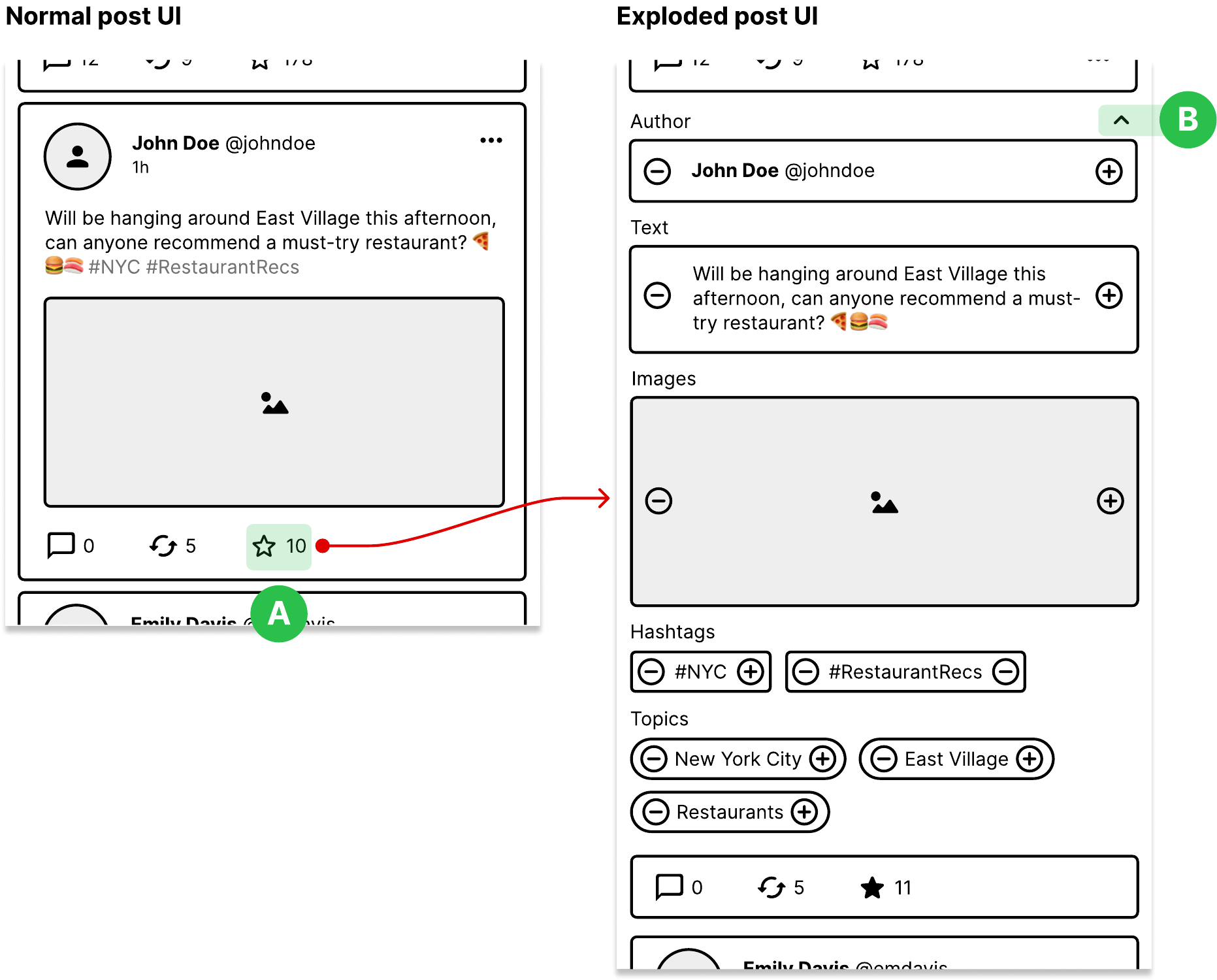}
    \caption{An example of the exploded view feed design with 5 features: author, text content, images, hashtags, and topics. \textbf{(A):} the exploded view is triggered as the user provides a generic signal---in this case, a ``like.'' \textbf{(B):} the exploded UI can be collapsed back into the normal UI using the caret.}
    \label{fig:exploded-views}
    \Description{Two mockups of feed UI screens side by side. The one on the left shows a post in a feed. The one on the right shows the same post but its UI is separated into smaller UIs, each containing a separate part of the original post. These parts include the post author, the post's text, and the hashtags included in the post.}
\end{figure*}

Once a user expresses a generic signal (e.g., like, reshare) on a post, the UI explodes out into individual features, such as the author, text descriptions, attached media, and hashtags (all features in our taxonomies---see Section \ref{s:taxonomies}). These features may be directly available in the post or algorithmically extracted---indeed, \textsc{topic} was a popular feature in our content-based taxonomy that needed to be inferred from posts (Fig. \ref{f:content-taxonomy-agg}). In Fig. \ref{fig:exploded-views}, we distinguish inferred features from extracted ones by rounding out their UI. The user can then engage in teaching the feed algorithm their preferences in this exploded view by selecting the features that they consider (ir)relevant. While we could further elicit detailed characteristics associated with those features, just like we did in our study's Miro board, we reduced this down to simple plus and minus buttons (indicating a positive or negative characteristic, respectively) to avoid overburdening the user. 

We ensure that ``explosion'' happens within the feed so that teaching can be done without leaving the feed, satisfying \textbf{D1}. Additionally, since the exploded view can occupy more space than the normal post UI, especially when there are numerous features available to teach with, we provide a convenient option for the user to collapse back the exploded view (see Fig. \ref{fig:exploded-views} (B)). The user can also simply scroll away. This simple dismissal of the exploded UI aligns with \textbf{D2}. Finally, even as direct manipulation interfaces affording immediate interaction, exploded UIs allow users to gradually accumulate preferences informed by what they view, satisfying \textbf{D5}. \looseness=-1

\subsection{Multi-Feed Curriculum Organization and Seeding}
\label{s:multi-feed}
A teaching language alone cannot close the teaching loop. Here, we propose a design in which assembling the teaching curriculum and evaluating the learner's performance is closely integrated with the teaching language via feed multiplicity. Key to this design is the use of curriculum organization to curate a multi-feed experience. 

\begin{figure*}[h]
    \centering
    \includegraphics[width=1\textwidth]{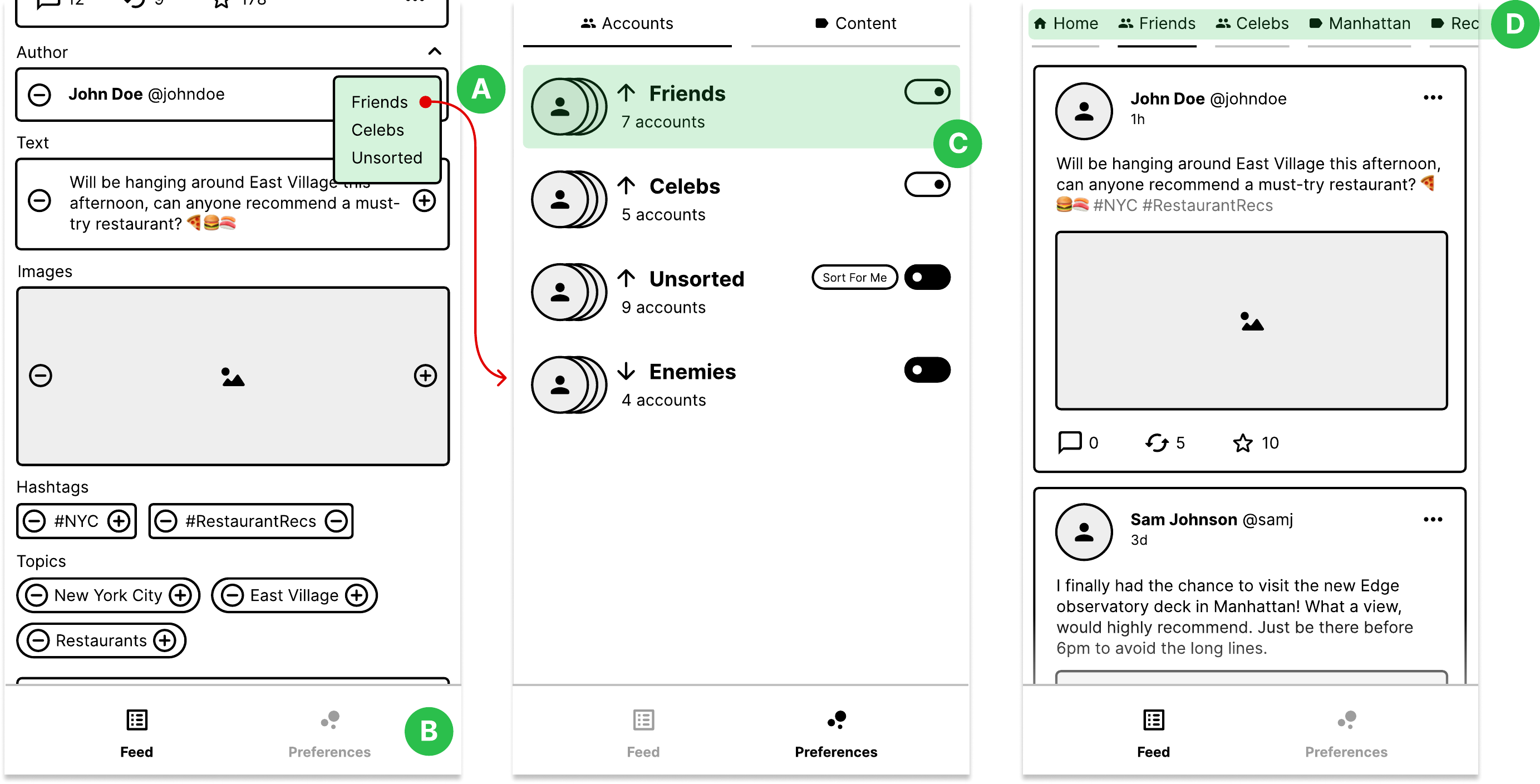}
    \caption{The teaching curriculum, depicted on the interface as ``Preferences,'' as a multi-feed experience. \textbf{(A):} the positive account-based feedback provided by the user in the exploded UI view prompts further organization from the user before the account is deposited into the curriculum. \textbf{(B):} tab to easily switch between viewing the feed and the curriculum. \textbf{(C):} the user's categorization of the account as ``Friend'' creates a new entry in the corresponding curriculum folder. Folder titles have arrows indicating whether the folder was created from positive or negative feedback. \textbf{(D):} feeds are formed from folders that users have toggled on. The posts on which users first provided feedback act as seeds for the folder feed.}
    \label{fig:multi-feed}
    \Description{Three mockups of feed UI screens side by side. The leftmost screen shows an exploded UI concept. The first screen leads to the middle screen, which shows 4 folders of organized content on a tab called Accounts. The other tab is called Content. The rightmost screen shows how each folder becomes a tab that contains a feed of its own. This is in addition to a Home feed that exists by default.}
\end{figure*}

As a user expresses preferences using a teaching language, those preferences are saved into curriculum ``folders.'' Due to the differing nature of account- and content-based preferences, we separate the two in Fig. \ref{fig:multi-feed}. A folder can then spawn a new feed that aims to provide more focused content that adheres to the folder's theme. The post on which the user first expressed preferences becomes a ``seed'' for the new feed to guide the recommendation of related content. This multi-feed experience serves as a way for users to evaluate the algorithmic learner's performance---a relevant and well-curated feed is a sign that the learner is effectively acting on taught preferences. Otherwise, the user can provide feedback to the learner through the in-feed teaching language, closing the teaching loop. \looseness=-1

If a user does not want a folder to form a feed, they can toggle it off in the curriculum; folders created from negative feedback are toggled off by default. Furthermore, as participants pointed out in Section \ref{s:archive-retrieval}, lightweight algorithmic interventions, such as suggesting folders for unorganized areas of the curriculum (see the ``Unsorted'' folder's Sort For Me option in Fig. \ref{fig:multi-feed}) can also aid with curriculum organization. 

\begin{figure*}[h]
    \centering
    \includegraphics[width=0.8\textwidth]{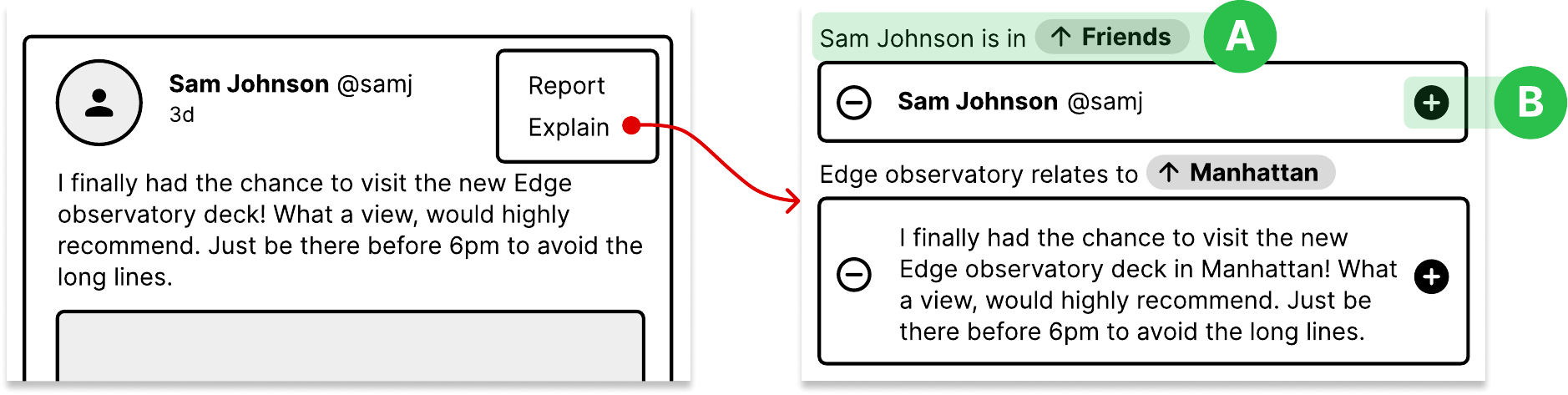}
    \caption{The teaching language (exploded UIs in this example) is invoked to scaffold learner evaluation once the ``Explain'' option is selected. \textbf{(A):} explanatory feature titles indicate relation to users' previously expressed preferences and links to an existing curriculum folder. \textbf{(B):} the learner pre-selects option(s) in the UI to show its preference prediction, which the user can correct by un-selecting if desired.}
    \label{fig:evaluation}
    \Description{Two mockups of feed UI screens side by side. The left screen shows a post that when selected, shows two options: Report and Explain. Selecting the Explain option will lead to the second screen, where the author and text within the post are in separate containers and indicate how they relate to previously expressed user preferences. These indications are pre-selection of the + button within the container and explanatory text in the container title that links to a curriculum folder.}
\end{figure*}

Given that user trust hinges on observed learner performance \cite{wall2019using, mosqueira2023human}, additional support may be added to further facilitate learner evaluation. One possible approach is to reuse already-familiar interaction patterns in the teaching language scaffold evaluation. Fig. \ref{fig:evaluation} offers an ``Explain'' option for posts recommended by the learner in a feed formed from a folder. Selecting that option would expand the post into an exploded UI with pre-selected preferences that the learner infers from existing ones. The titles of features then become explanations that link back to folders and other curriculum material used by the learner to infer that preference. With limited work in explanation and evaluation techniques in IMT \cite{teso2023leveraging}, employing the teaching language as an aid for evaluation suggests one approach to mend this gap.

This design's use of feed multiplicity embodies \textbf{D3}. Additionally, unlike the set-and-forget approach to creating custom feeds on Bluesky \cite{bluesky-feeds}, this design enables users to iteratively build, refine, and evaluate their feeds in the spirit of \textbf{D5}. This design can also be easily dismissed (\textbf{D2}): users can toggle feeds off from within the curriculum and the curriculum itself is located in another tab separate from regular feed activities. However, if the user does choose to engage with this design, the proximity to and reliance on an in-feed teaching language satisfies \textbf{D1}.

\subsection{Purposefully Finite Feeds with Natural Language Feedback}
The infinite scroll is a dominant design pattern in contemporary social media. Implementation-wise, this effect is achieved by loading posts in batches such that another batch of content is quickly available once the user reaches the end of the previous batch. What if we can repurpose this transition between batches into an opportunity for preference elicitation and reflection?

In this feed design, we explore what it means for feeds to be \textit{purposefully finite}. We split a feed into individual ``stacks'' of content; users can set the size (number of posts) of each stack. When the user reaches the end of a stack, they are presented with a teaching language in the form of a text input area in which they can specify, in natural language, any preferences to incorporate into future stacks. The combination of finite stacks and natural language feedback addresses two insights participants raised in our study. First, preferences may not be well-formed before consuming content---by allowing users to first view content and then reflect on what they viewed, users may be able to articulate their preferences with more precision. Second, the feed was used to both browse and search for content. An adjustable stack size allows a user to smoothly transition between the two consumption modes. With a low stack size and frequent feedback input, the feedback starts to resemble intent-driven search queries, while a high stack size brings the experience closer to that of a conventional infinitely scrolling feed. On top of all this, unstructured natural language allows users to specify nuances that may be difficult to communicate through structured means, such as UI buttons.  

\begin{figure*}[h]
    \centering
    \includegraphics[width=0.8\textwidth]{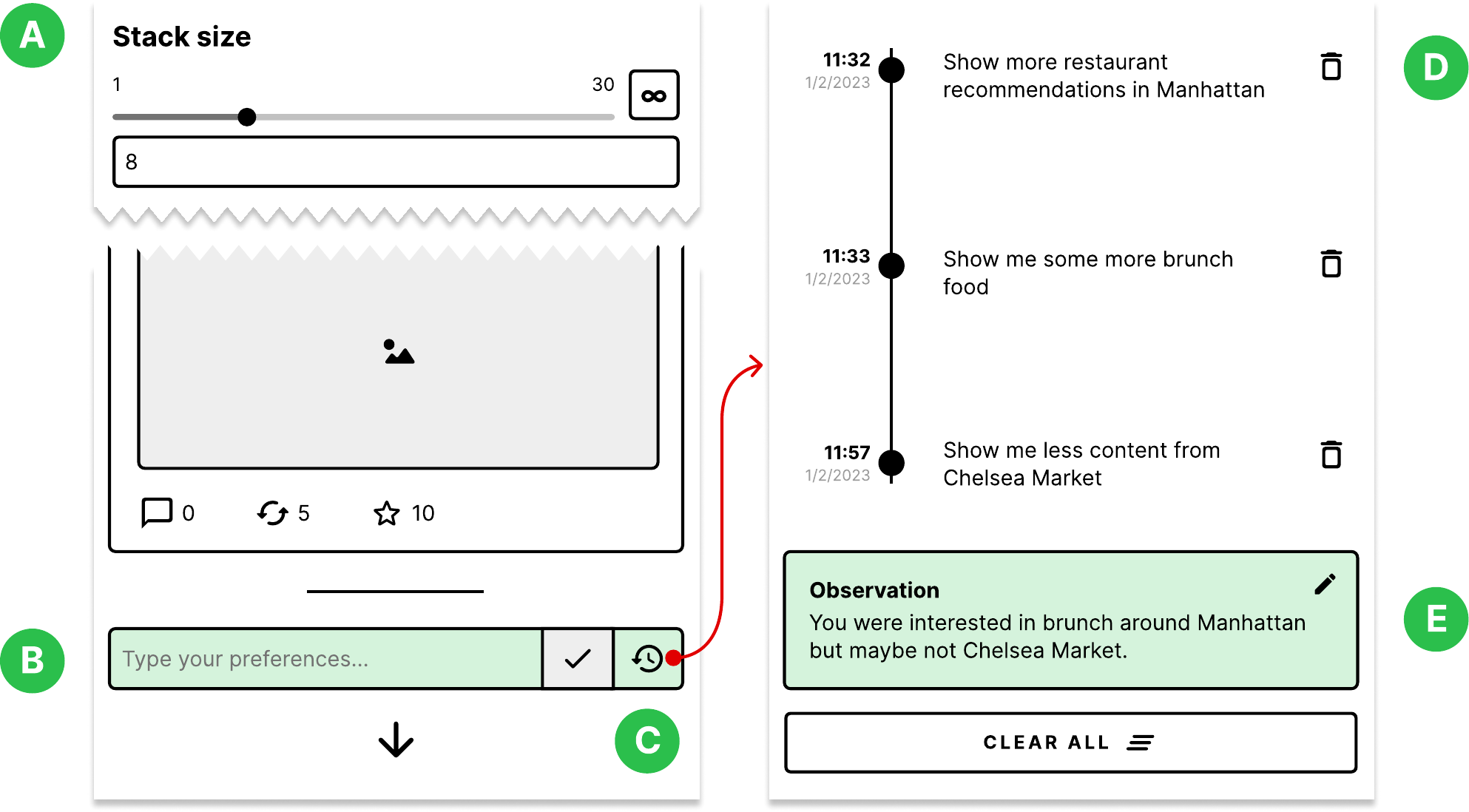}
    \caption{A stack of feed content followed by an opportunity for natural language feedback. \textbf{(A):} the user can set the size of the stack, or select the infinity option for an infinitely scrolling feed. \textbf{(B):} text box for unstructured natural language preference expression. \textbf{(C):} option to view the history of submitted preferences. \textbf{(D):} submitted preferences are cumulative and appear in a timeline. Users can remove individual preferences from the timeline or clear all of them using the option at the bottom of the page. \textbf{(E):} to aid user reflection, a synthesis of their preferences is generated and can be edited to guide future recommendations.}
    \label{fig:stacks}
    \Description{Two mockups of feed UI screens side by side. The left screen contains a slider at the top labelled Stack Size that is set to 8 but can be set from 1 to 30, with an additional button for infinity. The bottom of the screen is a text input bar containing placeholder text that reads ``Type your preferences'' with a checkmark button and a history button that leads into the second screen on the right. The second screen contains a vertical timeline showing 3 text inputs a user has provided within a period of 30 minutes and an Observation panel under the timeline that synthesizes the 3 inputs into a higher-level observation. At the very bottom of the screen, there is a Clear All button.}
\end{figure*}

The accumulation of natural language feedback also presents novel interaction opportunities. Users may use consecutive pieces of feedback to assemble a chain of preferences that can help them discover content with specific features and characteristics. Users can then exit from these more focused views by deleting preferences from their chain, or removing the entire chain to start afresh. Users' preferences can also be automatically summarized by the algorithmic learner into ``Observations'' (see Fig. \ref{fig:stacks}) are shown to the user for additional reflection. These observations can be edited by the user to guide future recommendations, similar to editable natural language user profiles in modern conversational recommender systems \cite{Friedman2023LeveragingLL}. 

This natural language feedback, when used in combination with the more structured feedback presented in Section \ref{s:exploded-ui}, results in an expressive set of teaching languages (\textbf{D4}). Like exploded UI views, it also operates in-feed, per \textbf{D1}. If the user does not wish to engage in this design, they may simply proceed to the next stack without providing feedback, or revert to an infinitely scrolling feed by setting the stack size to infinity, satisfying \textbf{D2}. Finally, by first accumulating feedback over time as they view content, users can refine system behavior over longer timescales, while still being afforded the ability to directly and immediately update the system's learned knowledge as preferences evolve (\textbf{D5}).

\label{s:patterns}

\section{Discussion}
\subsection{Ethical Implications}
\label{s:ethical-implications}
Our goal of enabling teachable social media feed experiences is to empower users in reclaiming agency and enriching self-expression. % through adapting an algorithmic learner to their personal tastes and preferences. 
While we hope that such a goal would bring positive change to social media systems, certain factors, if not carefully considered, may lead to the opposite outcome. 

\subsubsection{Issues around (hyper)personalization} Granular preferences captured through teachable feeds can lead to hyper-personalization---the use of deeply personal traits such as one's personality, motivations, and goals \cite{warshaw2015can, zhou2019trusting} to personalize experiences. Hyper-personalization may exacerbate privacy concerns many already have about social media platforms. In particular, users' lay knowledge of behavioral profiling for targeted advertisements has led to negative perceptions and distrust of platforms to responsibly store personal information \cite{barbosa2021design, ur2012smart}. Prior work has also warned about social pressures that can lead to over-sharing of hyper-personalized information on social media \cite{warshaw2015can}. %\remove[]{Additionally, hyper-personalization can deepen echo chambers by exposing users to homogenized views they already subscribe to.}

More broadly, John Cheney-Lippold coined the term ``dividuals'' \cite{cheney2017we} to highlight an inherent paradox of personalization on large-scale, data-driven platforms: one's ``personalized'' experiences online are never truly local to an individual, but are instead constantly shifting collections of data points---dividuals---aggregated across many users. That is, one is never an ``individual'' on these platforms, but is instead an algorithmically assembled profile of dividuals. Simpson et al. \cite{simpson2022tame} claim that this shifting assemblage results in an ``identity spiral'' in which users' online identities are always being revised and connected to other identities in uncontrolled and sometimes disagreeable ways. Enhanced personalization, including via teachable feeds, may perpetuate this spiral and dividualization as a whole. 

\subsubsection{Potential solutions} At the same time, teachable feeds have potential to alleviate some of these concerns. Many concerns around user privacy and targeting stem from the opacity with which algorithms operate---users are unaware of what kind of data is collected about them and where the data is transferred to after collection. By eliciting explicitly taught preferences, users can more clearly grasp what the algorithm has learned about them. From a platform's perspective, the improved granularity of user feedback may reduce the need for implicitly inferring certain preferences. Taken together, the reclamation of user agency may in fact aid platforms in reaching transparency ideals. Furthermore, by accumulating their own set of taught preferences, users can be their own architects of their individualized online identity rather than relying on an algorithm to assemble it from dividuals. While some forms of automated profiling may still persist, the nuanced experiences of the individuated user can now remain uniquely local, not simply dissolved into the dividual crowd.

We also hope that our proposed design patterns can pave the path for new forms of content moderation, such as detecting curriculum folders containing high concentrations of malicious content and natural language feedback that indicates intent to disseminate such content. 
This will hopefully disincentivize adversarial actors from using teachable feeds to spread harm (e.g., collaboratively plot disinformation campaigns \cite{starbird2019disinformation}). Similar methods can also be used to identify when users are exposed to increasingly homogeneous perspectives\footnote{We use ``increasingly homogeneous perspectives'' in this work but recognize the rich, yet inconclusive, body of empirical work investigating similar phenomena termed ``echo chambers'' and ``filter bubbles'' \cite{auxier2019factors, quattrociocchi2016echo, celis2019controlling, fletcher2021many, dubois2018echo, pariser2011filter}.} from their content, and when to introduce nudges to increase content diversity. Users can also use these moderation mechanisms to better reflect on their own content browsing habits, treating their role as a teacher \textit{as a learning experience}. Prior work in IMT showed that reflecting upon workplace calendar data throughout the teaching process surfaced new insights from personal data and informed future digital wellbeing goals \cite{jorke2023pearl}. We see room to further explore this in a social media context.

\subsection{Limitations and Future Work}
As with any sensitizing concept, our design patterns are meant to be iterated upon and refined. We showed that these patterns can embody the design principles derived from our study data, but we did not perform empirical user evaluations of our designs. Future work may integrate these designs into existing or new social media platforms and run user studies to gather feedback. Our design patterns were also applied to a microblogging-style platform similar to Mastodon or Twitter. Future work may explore translating them into a TikTok-style platform, as 1) the medium of video may cast additional considerations onto our proposed patterns, and 2) the user can only see one post at any given time, unlike the other feeds. Social media aside, other feed-based systems, such as news applications, may also benefit from our design patterns. Further research is needed to determine if or how they should be applied in domains outside of social media.

Teachable feeds also present exciting opportunities for shared, collaborative IMT experiences. For example, a user who has a curriculum folder for celebrities they are interested in seeing more of in their feed may share that folder with a friend with similar celebrity interests. Users living in the same town or city may share curricula with local events or news. More broadly, users who share their established curricula assembled over time can help those with less experience as a teacher in IMT jumpstart the IMT process. Collaborative IMT can thus be of particular interest to future work aiming to expand existing strategies to support novice teachers, strategies that are currently limited to showing push notifications of expert teachers' best practices using Wizard-of-Oz methods \cite{wall2019using}.

Additionally, the platforms used in our study were more different than they were similar---users opened their feeds for different purposes, followed different accounts, and were exposed to different feed affordances. While this enabled us to create a broader mapping of our design space and witness shifts in taxonomies across the platforms, it did not allow us to isolate any particular variables. The launch of Threads \cite{threads} in July 2023 presents a rare opportunity to address this. Because Instagram and Threads accounts are automatically linked, one may follow the same accounts on both platforms but find themselves in two completely distinct user experiences. Future work can use this as a case study for how variations in feed affordances can impact content perception and preferences, thus shaping the ``culture'' of a platform. 

Finally, even though we proposed teachable feeds with users' best interests in mind, we acknowledge that social media platforms may not always act in those interests en route to achieving their financial goals \cite{bruckman2022should}. We do, however, see potential avenues for incentive alignment and adoption of our proposed designs. Currently, the dominant financial model for platforms revolves around maximizing ad revenue by engagement optimizing techniques to prolong users' ad exposure. We encourage future work to investigate how to \textit{increase ad quality} using feedback willingly provided by users in the teaching loop to reduce reliance on \textit{ad quantity} and \textit{ad exposure time}. This way, ad targeting can be made more transparent, as users have a record of what they taught to their feed. Irrelevant ads, which clutter many of today's platforms, may also be reduced upon platforms receiving higher-quality user feedback. Platforms may, and have already started to, explore new financial models.
For example, Mastodon has adopted a non-profit financial model, choosing ``not [to] implement any monetization strategies in the software'' {\cite{mastodon-docs}} and instead funding ongoing development with grants and crowd-sourced funds {\cite{investopedia-mastodon}}. This lack of commercial involvement means that Mastodon can better align with user interests but constrains development resources to build experiences that are polished and compelling enough for mass-market adoption {\cite{bruckman2022should}}. Public funding may be crucial in the success of these non-profit models.
As such, policymakers may see more opportunities to engage with this space to advocate for users while fostering financially sustainable platforms.

\section{Conclusion}
An architect may design a home, but it is ultimately up to the home's occupants to decorate the space to reflect their unique tastes and preferences. Today, social media feeds act as digital architects of personalized social spaces, but users often lack such decorative agency. \looseness=-1

In this work, we explored the idea of teachable social media feed experiences for agential, personalized feed curation. To do so, we conducted a study with 24 social media users across Instagram, Mastodon, TikTok, and Twitter to elicit key signals they used to determine the value of posts in their feeds. We found that users evaluated content in multi-faceted and nuanced ways that cannot be fully captured by affordances on current platforms. To enable users to better ``teach'' an algorithm their preferences, we offered five IMT-inspired five design principles for teachable feed experiences, informed by findings from our study. We then embodied these principles in three feed designs to inspire future efforts on integrating teachable feeds into real-world social media systems.

Altogether, our contributions lay the groundwork for continued exploration of teachable feed experiences. We hope this exploration can empower users and platforms to craft more comforting and expressive digital homes.
\bibliographystyle{ACM-Reference-Format}
\bibliography{refs}

\end{document}